\documentclass[sigplan,10pt]{acmart}
\renewcommand\footnotetextcopyrightpermission[1]{}
\makeatletter
\def\@ACM@checkaffil{%
    \if@ACM@instpresent\else
    \ClassWarningNoLine{\@classname}{No institution present for an affiliation}%
    \fi
    \if@ACM@citypresent\else
    \ClassWarningNoLine{\@classname}{No city present for an affiliation}%
    \fi
    \if@ACM@countrypresent\else
        \ClassWarningNoLine{\@classname}{No country present for an affiliation}%
    \fi
}
\makeatother

\usepackage{tikz}
\usepackage{amsmath}

\usepackage{filecontents}

\usepackage[normalem]{ulem}
\usepackage{etex}
\usepackage{tikz}
\usepackage{amsmath}

\usepackage{filecontents}
\usepackage{color}
\usepackage{xcolor}
\usepackage[compact]{titlesec}
\usepackage{xspace}
\usepackage{enumitem}
\usepackage{mfirstuc}
\usepackage{cleveref}
\crefname{section}{§}{§§}
\usepackage{caption}
\usepackage[labelformat=simple]{subcaption}

\usepackage{circledsteps}

\usepackage{booktabs}
\usepackage{multirow}
\usepackage{pifont}
\usepackage{tablefootnote}
\usepackage{makecell}
\usepackage{bookmark}
\hypersetup{
  colorlinks=true,
  linkcolor=blue!90!red,
  citecolor=magenta
}
\usepackage[ruled,vlined,linesnumbered]{algorithm2e}
\makeatletter
\renewcommand{\sectionautorefname}{\S\@gobble}
\renewcommand{\subsectionautorefname}{\S\@gobble}
\renewcommand{\subsubsectionautorefname}{\S\@gobble}

\setenumerate[1]{itemsep=0pt,partopsep=0pt,parsep=\parskip,topsep=5pt}
\setitemize[1]{itemsep=0pt,partopsep=0pt,parsep=\parskip,topsep=5pt}

\newcommand{\sys}{S\small INLK}
\newcommand{\chm}{CXL-HM}
\newcommand{\hm}{HM}
\newcommand{\sart}{S-ART}
\newcommand{\smass}{S-Masstree}
\newcommand{\massl}{Mass-L}

\newcommand{\codeword}[1]{\textcolor{black}{$\mathsf{#1}$}}

\usepackage[misc]{ifsym}
\usepackage{bbding}
\usepackage{breakurl}
\usepackage{wasysym}
\usepackage{tabularx}
\usepackage{wrapfig}
\usepackage{textcomp}
\usepackage{listings}
\usepackage{pifont}
\usepackage[many]{tcolorbox}
\usepackage{float}
\usepackage{hyperref}
\newfloat{lstfloat}{t}{lop}
\floatname{lstfloat}{Listing}

\definecolor{blue-violet}{rgb}{0.54, 0.17, 0.89}
\definecolor{red-violet}{rgb}{0.78, 0.08, 0.52}
\definecolor{darkgreen}{rgb}{0.0, 0.2, 0.13}
\definecolor{islamicgreen}{rgb}{0.0, 0.56, 0.0}
\definecolor{officegreen}{rgb}{0.0, 0.5, 0.0}
\definecolor{hotmagenta}{rgb}{1.0, 0.11, 0.81}
\definecolor{codegreen}{rgb}{0,0.6,0}
\definecolor{codegray}{rgb}{0.5,0.5,0.5}
\definecolor{codepurple}{rgb}{0.58,0,0.82}
\definecolor{backcolour}{rgb}{1,1,1}
\definecolor{codeblue}{rgb}{0.25,0.5,0.5}
\definecolor{codemaroon}{rgb}{0.5,0,0}

\begin{document}

\pagenumbering{arabic}

\newcommand{\titlebody}{Optimizing Tree-structure Indexes for CXL-based Heterogeneous Memory}

\title{{\titlebody} with S{\fontsize{15}{17}\selectfont INLK}}

\author{\rm Haoru Zhao,\; Mingkai Dong\textsuperscript{\Envelope},\; Fangnuo Wu,\; Haibo Chen\\
{\normalsize {Institute of Parallel and Distributed Systems, Shanghai Jiao Tong University}} \\
}

\sloppy

\begin{abstract}

On heterogeneous memory ({\hm}) where fast memory (i.e., CPU-attached DRAM) and slow memory (e.g., remote NUMA memory, RDMA-connected memory, Persistent Memory (PM)) coexist, optimizing the placement of tree-structure indexes (e.g., B+tree) is crucial to achieving high performance while enjoying memory expansion.  
Nowadays, CXL-based heterogeneous memory ({\chm}) is emerging due to its high efficiency and memory semantics.
Prior tree-structure index placement schemes for {\hm} cannot effectively boost performance on {\chm}, as they fail to adapt to the changes in hardware characteristics and semantics.
Additionally, existing {\chm} page-level data placement schemes are not efficient for tree-structure indexes due to the granularity mismatch between the tree nodes and the page.

In this paper, we argue for a \emph{CXL native, tree-structure aware} data placement scheme to optimize tree-structure indexes on {\chm}.
Our key insight is that \emph{the placement of tree-structure indexes on {\chm} should match the tree's inherent characteristics with {\chm} features.}
We present {\sys}, a tree-structure aware, node-grained data placement scheme for tree-structure indexes on {\chm}.
With {\sys}, developers can easily adapt existing tree-structure indexes to {\chm}.
We have integrated the B+tree and radix tree with {\sys} to demonstrate its effectiveness.
Evaluations show that {\sys} improves throughput by up to 71\% and reduces P99 latency by up to 81\% compared with state-of-the-art data placement schemes (e.g., MEMTIS) and {\hm}-optimized tree-structure indexes in YCSB and real-world workloads.
 
\end{abstract}

\settopmatter{printfolios=true,printacmref=false}
\maketitle
\pagestyle{plain}

\renewcommand*{\thefootnote}{{\Envelope}}
\footnotetext[1]{ Corresponding author: Mingkai Dong (\burl{mingkaidong@sjtu.edu.cn}).}
\renewcommand{\thefootnote}{\arabic{footnote}}
\setcounter{footnote}{0}

\thispagestyle{empty}

\section{Introduction}%
\label{sec:intro}

In-memory tree-structure indexes~\cite{mao2012cache, wu2019wormhole, zeitak2021cuckoo, binna2018hot, kim2010fast,rao2000making,levandoski2013bw,shanny2022occualizer,wang2018building,leis2013adaptive,leis2016art} (e.g., B+tree) are fundamental in various domains, such as databases~\cite{tu2013speedy,zhang2016reducing,HyPer,MemSQL,SAP_HANA,TimesTen,VoltDB}, storage engines~\cite{lepers2019kvell,lim2011silt} and big data analytics~\cite{armbrust2015spark,korber2021index,druid,flink,elasticsearch}.
Their efficiency is crucial for overall system performance~\cite{zeitak2021cuckoo,binna2018hot,zhang2016reducing,kocberber2013meet}.
With the increasing data scale in datacenters, the memory demand of tree-structure indexes is growing rapidly~\cite{anneser2022adaptive,zhang2020order,atikoglu2012workload,lim2014mica,zeitak2021cuckoo,raybuck2021hemem,zhang2015memory}.
In order to expand memory capacity and enhance memory elasticity, heterogeneous memory (HM) has emerged.
Many efforts~\cite{metreveli2012cphash,calciu2013using,calciu2017black,ziegler2019designing,wang2022sherman,luo2023smart,luo2024chime,lu2024dex,li2023rolex,yang2015nv,oukid2016fptree,se2017wort,zhou2019dptree,chen2020utree,liu2020lb,zhang2022nbtree} have been made to build tree-structure indexes on {\hm} to leverage the large memory capacity while mitigating the slow memory's impact on performance.

Nowadays, Compute Express Link (CXL)~\cite{cxl_homepage} becomes a promising {\hm} solution to expand memory capacity.
For example, Samsung CMM-B~\cite{samsung_cmmb} scales memory capacity up to 24\,TB with CXL memory pool.
Due to CXL's cache-coherent memory semantics (i.e., CXL provides byte-addressable memory accessed in a cache-coherent way in the same address space as CPU-attached DRAM, i.e., fast memory), large tree-structure indexes designed for fast memory can easily adapt to {\chm}, where fast and CXL-attached memory coexist.

However, placing these tree-structure indexes designed for fast memory directly onto {\chm} results in poor performance, as there is still a performance gap between CXL-attached memory and fast memory (latency \textasciitilde2$\times$, bandwidth \textasciitilde60\%, as shown in \autoref{fig:mlc}).
For instance, directly placing 75\% of the tree's memory on CXL-attached memory can lead to a performance drop of \textasciitilde70\% (\autoref{fig:vary_cxl_percentage}). 
Therefore, it is vital to optimize the placement of tree-structure indexes on {\chm} to reduce performance degradation while taking advantage of the increased memory capacity.

Previous optimizations for tree-structure index placement on {\hm} (e.g., NUMA, RDMA, PM) are inefficient for {\chm}.
Replication, partitioning, and delegation methods used in NUMA-optimized indexes~\cite{calciu2013using,calciu2017black,metreveli2012cphash} are not suitable for {\chm} because CXL-attached memory lacks local processors like NUMA nodes.
Indexes optimized for RDMA~\cite{ziegler2019designing,wang2022sherman,luo2023smart,luo2024chime,lu2024dex} maintain a cache for accessed data in fast memory.
However, the narrowed performance gap between slow and fast memory in {\chm} makes cache maintenance overhead outweigh the benefits of caching.
In indexes optimized for PM~\cite{yang2015nv,oukid2016fptree,se2017wort,chen2020utree,liu2020lb,zhang2022nbtree,zhou2019dptree}, placing the internal nodes of trees in fast memory is commonly used to mitigate slow memory's limited write bandwidth and higher latency.
For {\chm}, this static approach is insufficient because it neglects the capacity limit of fast memory, leading to fast memory exhaustion or underutilization, and fails to adapt to dynamic workloads.

Moreover, applying existing {\chm} data placement schemes~\cite{maruf2023tpp,lee2023memtis,xiang2024nomad,vuppalapati2024tiered,zhou2024neomemhardwaresoftwarecodesigncxlnative} to tree-structure indexes is also inefficient.
As these schemes manage data at page granularity while tree nodes are typically smaller, this granularity mismatch leads to inaccurate hotness detection, increased migration overhead, and suboptimal data placement.

Our key insight is that \emph{the placement of tree-structure indexes on {\chm} should match the tree's inherent characteristics with the features of {\chm}.}
We identify three key characteristics of trees:
(1) \emph{node granularity}: node is the unit of data usage and management in trees;
(2) \emph{layered hotness}: nodes in upper levels are accessed more frequently (i.e., hot) than those in lower levels;
(3) \emph{path-style access}: the access granularity of trees is the path (i.e., all nodes from the root to a leaf
\footnote{We mainly focus on tree-structure indexes that store values in leaves, the most common form of such indexes~\cite{leis2013adaptive}.}
), and nodes along paths to hot leaf nodes (i.e., hot path) contribute to considerable memory access.

Therefore, on {\chm} we should focus on:
(1) Match the node granularity with {\chm}'s cache-coherent memory semantics.
Instead of caching, we should store hot nodes in fast memory and less frequently accessed (i.e., cold) nodes in slow memory, track node hotness, and migrate nodes between fast and slow memory based on the hotness to adapt to dynamic workloads.
(2) Match the layered hotness and path-style access with the performance differences between fast and slow memory by prioritizing upper-level nodes (i.e., \emph{layer principle}) and hot paths (i.e., \emph{path principle}) in fast memory.

We design {\sys}, a node-grained, tree-structure aware data placement scheme to optimize tree-structure indexes on {\chm}.
First, {\sys} introduces \emph{leaf-centric access tracking}, capturing the hotness of all access paths while minimizing interference with tree operations by only tracking leaf nodes.
Second, {\sys} incorporates \emph{structure-aware migration} which considers node relationships instead of migrating each node independently to keep upper-level nodes in fast memory and achieve the quick promotion (i.e., moving to fast memory) of the entire hot path.
Finally, {\sys} proposes \emph{hyper watermark mechanism} that holistically coordinates system behaviors (i.e., node allocation and migration) based on real-time fast memory usage rather than static thresholds, thereby ensuring stable high performance for indexes.

To demonstrate {\sys}'s effectiveness and generality, we integrate {\sys} with two widely used tree-structure indexes---B+ tree~\cite{mao2012cache} and radix tree~\cite{leis2013adaptive,leis2016art}---with less than 3\% trees' internal code modification.
Evaluation on a real CXL platform with synthetic workloads shows that {\sys} achieves up to 133\% higher throughput and 67\% lower P99 latency than the SOTA data placement scheme MEMTIS~\cite{lee2023memtis}.
For YCSB~\cite{YCSB2010}, {\sys} achieves up to 71\% and 60\% higher throughput than MEMTIS and optimized {\hm} index (optimized PACTree~\cite{kim2021pactree}), respectively.
For real-world workloads~\cite{alibaba_block_traces}, {\sys} achieves up to 66\% higher throughput and 81\% lower P99 latency than MEMTIS, with 44\% higher throughput and 63\% lower P99 latency than optimized PACTree.

In summary, we make the following contributions:
\begin{itemize}[leftmargin=1em,topsep=-0.14pt]
    \item \textbf{Analyses.} We thoroughly analyze existing solutions for the placement of tree-structure index on {\chm} and propose that the key to efficient placement lies in matching the tree's inherent characteristics with {\chm} features.
    \item \textbf{Design.} We design {\sys}, an efficient and generic data placement scheme for tree-structure indexes on {\chm}.
    \item \textbf{Evaluation.} We integrate {\sys} with B+ tree and radix tree. Evaluation on a real CXL platform with various workloads demonstrates the advantages of {\sys}. 
\end{itemize}

\section{Background and Motivation}%
\label{sec:back}

\subsection{CXL-based Heterogeneous Memory}

\begin{figure}[t]
    \centering
    \subfloat[CXL 1.0/1.1]{
        \includegraphics[width=0.245\linewidth]{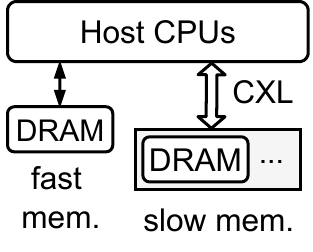}
        \vspace{-7.5pt}
        \label{fig:cxl-expansion}}
    \subfloat[CXL 2.0/3.0]{
        \includegraphics[width=0.35\linewidth]{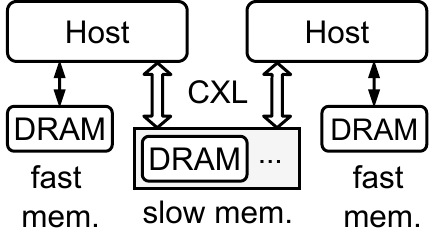}
        \vspace{-6pt}
        \label{fig:cxl-pooling}}
    \subfloat[Performance]{
        \includegraphics[width=0.34\linewidth]{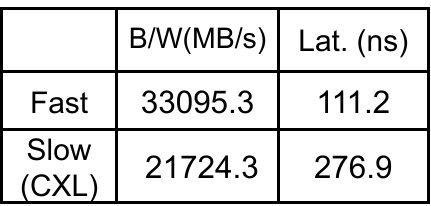}
        \vspace{2pt}
        \label{fig:mlc}}
    \vspace{-10pt}
    \caption{\textbf{CXL-based heterogeneous memory system.} \emph{(a,b) System architecture. (c) Comparison of fast and slow memory latency and bandwidth tested by Intel MLC~\cite{mlc} on our CXL1.1 platform.}}
    \label{fig:cxl-system}
\end{figure}

CXL~\cite{cxl_homepage, sharma2024introduction} is an emerging interconnect technology notable for its memory expansion capability, cache-coherent memory semantics, and high performance.
As illustrated in \autoref{fig:cxl-expansion}, {\chm} with CXL\,1.0/\,1.1~\cite{cxl_1.0} allows host CPUs to access CXL-attached memory with cache-coherent load/store.
{\chm} with CXL\,2.0/3.0~\cite{cxl_2.0,cxl_3.0} (\autoref{fig:cxl-pooling}) further allows the hosts to expand the memory by CXL memory pool, enhancing memory expansion's flexibility.
As {\chm} provides a unified physical address space of fast memory and CXL-attached memory, existing tree-structure indexes for fast memory can be effortlessly migrated to {\chm}.

\subsection{Tree-structure Indexes on {\chm}}
\label{sec:directly-cxl-index}

\begin{figure}[t]
    \centering
        \includegraphics[width=\linewidth]{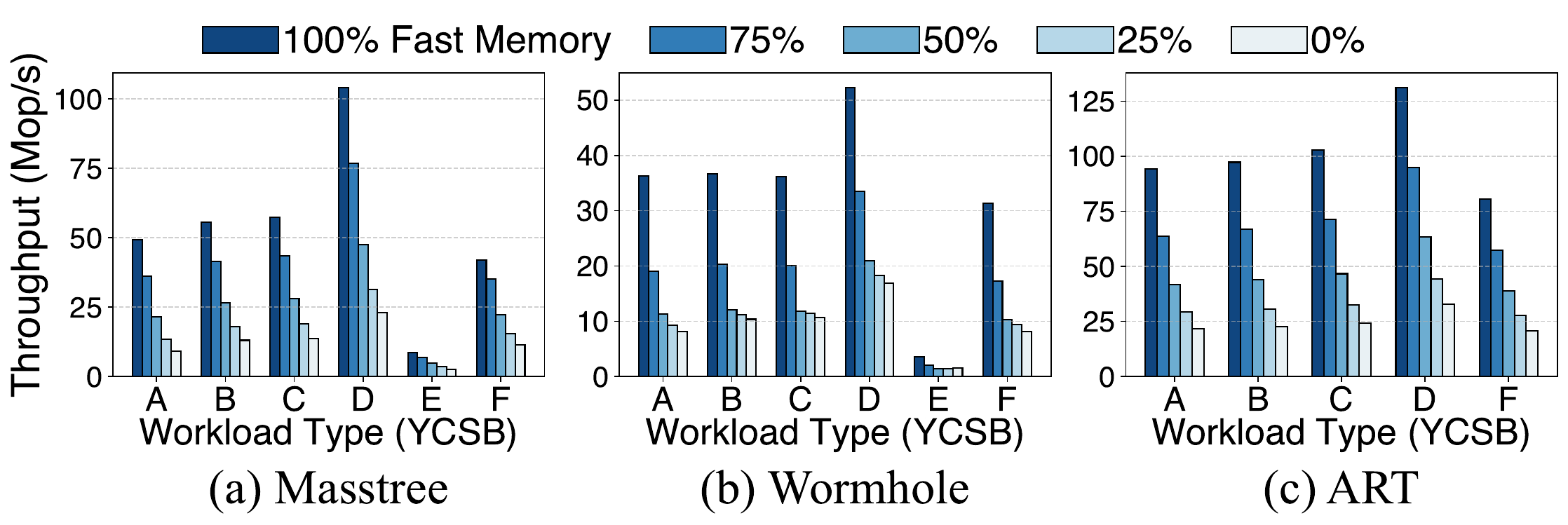}
        \vspace{-22pt}
        \caption{\textbf{Throughput (tput.) of indexes with different fast memory usage ratios in YCSB~\cite{YCSB2010} benchmark.} 
        }
    \label{fig:vary_cxl_percentage}
    \vspace{-8pt}
\end{figure}

\autoref{fig:mlc} shows that there is still a performance gap between fast and slow memory in {\chm}: slow memory has higher latency (\textasciitilde2$\times$) and lower bandwidth (\textasciitilde60\%) than fast memory on our CXL1.1 platform.
Moreover, this gap will be widened with the introduction of CXL-switch since CXL2.0~\cite{cxl_switch}.
Thus, directly migrating tree-structure indexes designed for fast memory to {\chm} leads to poor performance.
We illustrate this via Masstree~\cite{mao2012cache}, Wormhole~\cite{wu2019wormhole}, and adaptive radix tree (ART)~\cite{leis2013adaptive,leis2016art}, three typical tree-structure indexes.
The setup of the following experiments is the same as our evaluation (\autoref{sec:eval-setup}).
As shown in \autoref{fig:vary_cxl_percentage}\footnote{We don't test workload E for ART because its source code does not implement the scan operation.}, when employing the default weighted memory allocation policy~\cite{linux_weighted_interleave}, the performance of all three indexes drops by about 70\% with 25\% fast memory usage compared to 100\% fast memory across all workloads.
Therefore, \emph{optimizing the placement of tree-structure indexes on {\chm} is essential to achieve better performance.}

\subsection{Issues with Optimized HM-Indexes on {\chm}}
\label{sub:fast_memory_internode}

\begin{figure}[t]
    \centering
    \subfloat[Throughput]{
        \includegraphics[width=0.3\textwidth]{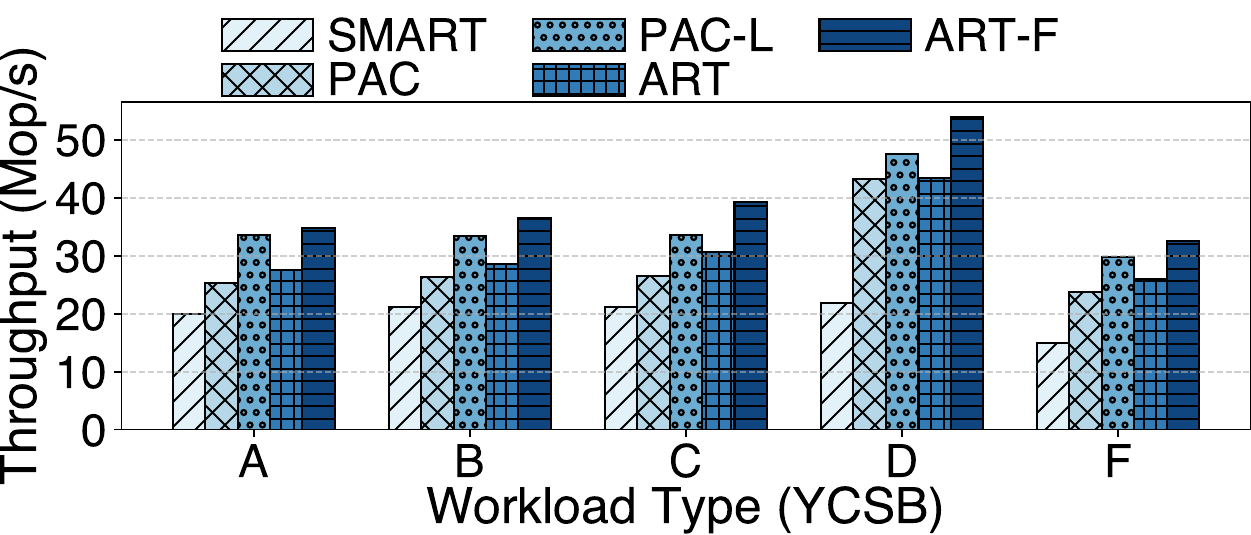}
        \vspace{-4pt}
        \label{fig:smart_vs_art_vs_pac_tput}
    }
    \subfloat[P99 Latency]{
        \includegraphics[width=0.165\textwidth]{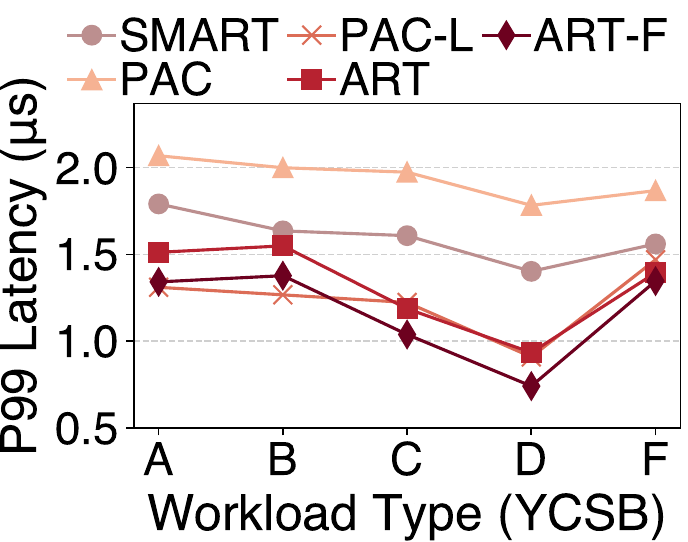}
        \vspace{-4pt}
        \label{fig:smart_vs_art_vs_pac_lat}
    }
    \vspace{-10pt}
    \caption{\textbf{Comparison of SMART, PACTree, and ART on {\chm}.} 
    {\emph{PAC stands for PACTree. 
    All indexes have a fast memory usage ratio of 20\%, except ART-F at 40\%.
    PAC, ART, and ART-F use the same memory allocation policy as \autoref{fig:vary_cxl_percentage}.} 
    }
    }
    \label{fig:smart_vs_art_vs_pac}
    \vspace{-5pt}
\end{figure}

\begin{figure}[t]
    \centering
    \begin{minipage}[b]{0.225\textwidth}
        \centering
        \includegraphics[width=\textwidth]{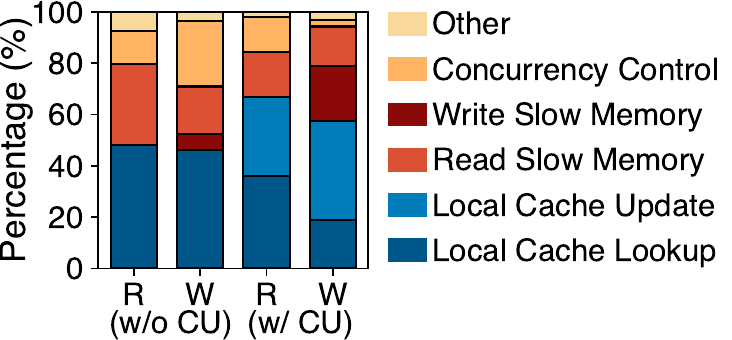}
        \vspace{-18pt}
        \subcaption{SMART}
        \label{fig:smart_breakdown}
    \end{minipage}
    \begin{minipage}[b]{0.245\textwidth}
        \centering
        \includegraphics[width=\textwidth]{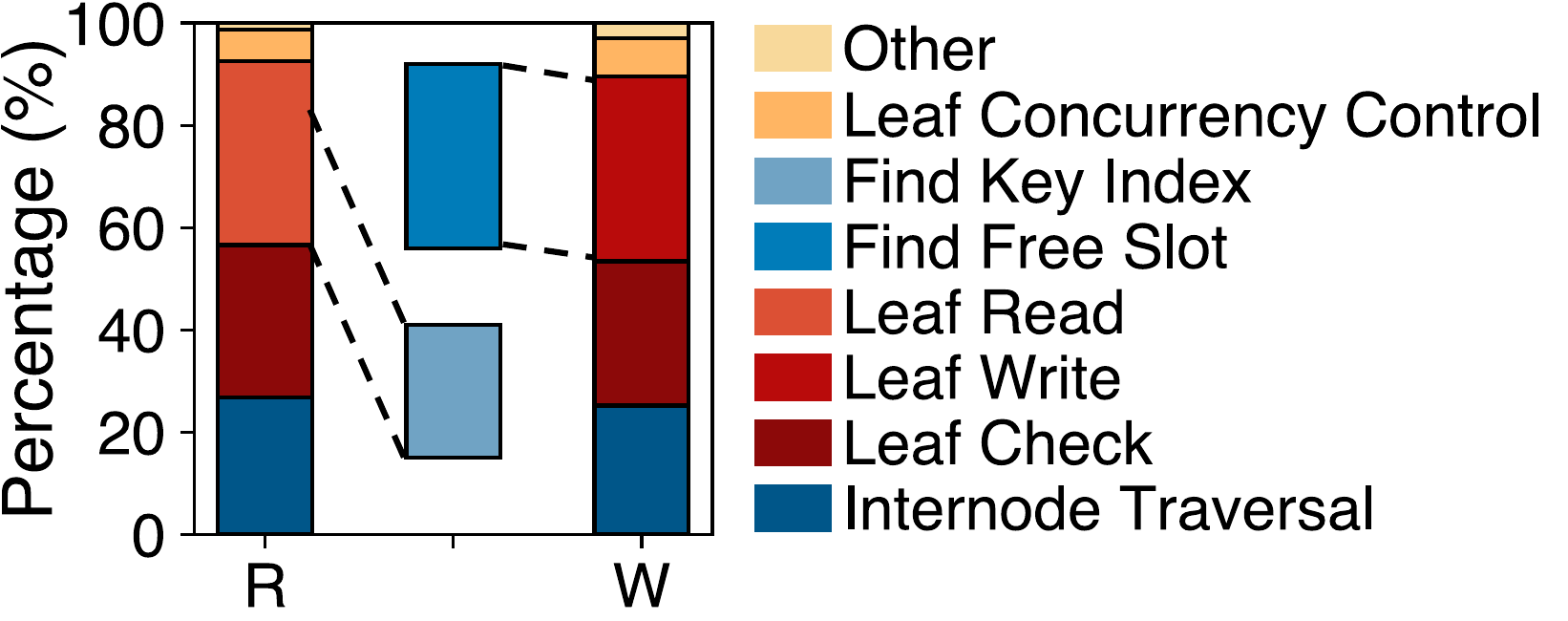}
        \vspace{-15pt}
        \subcaption{PACTree}
        \label{fig:pac_breakdown}
    \end{minipage}
    \vspace{-22pt}
    \caption{\textbf{Performance breakdown per operation type.} 
    {\emph{(w/o CU)\&(w/ CU) means without/with local cache update.}}}
    \label{fig:smart_breakdown_fig}
    \vspace{-8pt}
\end{figure}

In this section, we study existing tree-structure indexes optimized for other types of {\hm} (e.g., NUMA, RDMA-based {\hm}, PM-based {\hm}) and data placement schemes for {\chm} to demonstrate that:
\begin{itemize}[leftmargin=1em,topsep=1pt]
\item Prior optimizations for tree-structure index placement on {\hm} are either impractical or inefficient on {\chm} due to {\chm}'s unique hardware characteristics and semantics. We compare the original ART, the RDMA-optimized ART (SMART~\cite{luo2023smart}), and the PM-optimized ART (PACTree~\cite{kim2021pactree}) using YCSB.
\autoref{fig:smart_vs_art_vs_pac} shows that \emph{the original ART outperforms SMART and PACTree on {\chm}}.
\item Applying existing {\chm} data placement schemes to tree-structure indexes is also inefficient due to granularity mismatch between tree nodes and pages.
\end{itemize}

\noindent
\textbf{NUMA-Index.}
Indexes optimized for NUMA reduce cross-node communication and minimize accesses to remote memory through replication~\cite{calciu2017black,qu2024wasp,daly2018numask}, partitioning~\cite{metreveli2012cphash}, or delegation~\cite{calciu2013using,strati2019adaptive}.
These methods are not applicable to {\chm} because \emph{CXL-attached memory lacks local processors for data processing, unlike the memory in the NUMA node}.

\noindent
\textbf{RDMA-Index Case Study (SMART).}
SMART optimizes ART's data placement by maintaining a fast memory cache for accessed internal nodes (a.k.a., local cache), which is a common optimization in RDMA-Indexes~\cite{wang2022sherman,luo2024chime,luo2023smart,li2023rolex}.
\autoref{fig:smart_vs_art_vs_pac_tput} shows that SMART's throughput is only 50\%--74\% of ART\footnote{The congestion control design of SMART (RDWC) is disabled because it causes about a 15\% performance drop when there is no congestion.}.
Our profiling (\autoref{fig:smart_breakdown}) reveals that the primary overhead (\textasciitilde50\%) in SMART comes from the local cache.
\emph{Since the latency gap between CXL-attached memory and fast memory is much smaller than that of RDMA-connected memory~\cite{lu2024scythe}, the costs of cache maintenance and querying outweigh the benefits of reducing slow memory accesses.}
Specifically, while the local cache significantly reduces access to slow memory (nearly all internal node accesses hit the cache in the Zipfian read/update-only workloads), querying the local cache incurs more overhead than direct tree traversal due to the additional cache miss checks and eviction management.

\noindent
\textbf{PM-Index Case Study (PACTree).}
PACTree optimizes ART's data placement by using fat (64 entries), unordered leaf nodes, which is a common optimization~\cite{chen2020utree,ma2021roart,zhang2022plin,yang2015nv,zhang2022nbtree,chen2015persistent,oukid2016fptree,liu2020lb} in PM-indexes, along with asynchronous tree structural modification (a.k.a., async SMO).
\autoref{fig:smart_vs_art_vs_pac} shows that PACTree's throughput is only 90\% of ART, and its tail latency is 1.3\textasciitilde1.9$\times$ of ART.
Our breakdown (\autoref{fig:pac_breakdown}) indicates that the main source of performance overhead (\textasciitilde61\%) is the fat, unordered leaf and async SMO.
\emph{These optimizations incur additional overhead in {\chm} because CXL-attached DRAM does not suffer from the read-write asymmetry present in PM~\cite{raybuck2021hemem,song2023prism}, and the high allocation costs associated with persistence~\cite{ma2021roart,kim2021pactree}.}
Although fat, unordered leaf nodes reduce write amplification and allocation overhead, they impose significant search overhead, with nearly 99\% of the time in the leaf write operation spent finding a free position.
Additionally, while async SMO decouples the write overhead of slow memory from the critical path, it introduces substantial overhead---particularly harming tail latency---through leaf checking, which ensures the correctness of async SMO.

Another common placement optimization in PM-indexes is to place only leaves in slow memory and keep internal nodes in fast memory (a.k.a., fast memory internode)~\cite{yang2015nv,oukid2016fptree,zhou2019dptree,liu2020lb,chen2020utree,zhang2022nbtree}.
We apply this technique to PACTree (PAC-L in \autoref{fig:smart_vs_art_vs_pac}).
Although it can improve PACTree's performance on {\chm}, this static placement strategy is insufficient because: 
(1) It neglects fast memory capacity, which can result in fast memory underutilization or exhaustion. 
For example, ART outperforms PAC-L when fast memory capacity is larger (ART-F in \autoref{fig:smart_vs_art_vs_pac}), and PAC-L's performance drops by \textasciitilde17\% when fast memory is full.
(2) It ignores workload characteristics, requiring access to the leaf in slow memory for each request.

\begin{figure}[t]
    \centering
    \subfloat[Masstree]{
        \centering
        \includegraphics[width=0.43\linewidth]{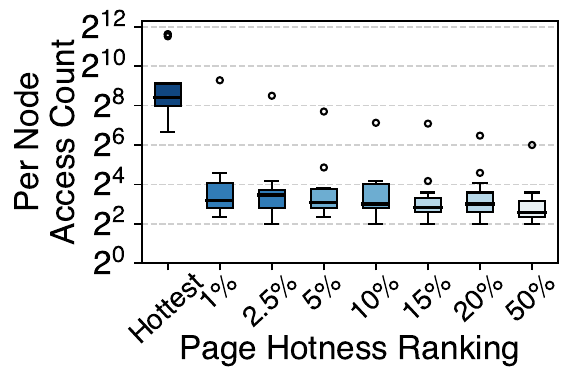}
        \vspace{-8pt}
        \label{fig:masstree_hotness_distribution}
    }
    \subfloat[ART]{
        \centering
        \includegraphics[width=0.397\linewidth]{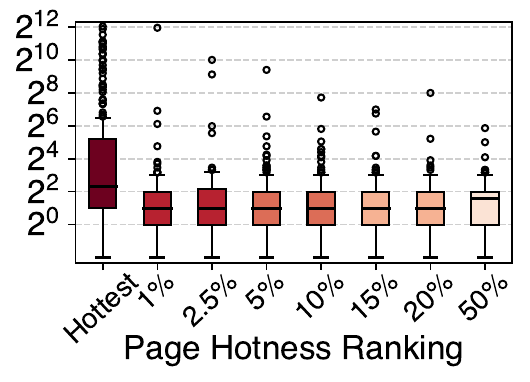}
        \vspace{-8pt}
        \label{fig:art_hotness_distribution}
    }
     \vspace{-10pt}
    \caption{\textbf{Node access count distribution across pages of varying hotness.} {\emph{We analyze distribution of node access counts in the page with the highest access frequency (Hottest) and those ranked top 1\%--50\% by access frequency.}}}
    \label{fig:hotness_distribution}
    \vspace{-6pt}
\end{figure}

\noindent
\textbf{Applying {\chm} Data Placement Schemes.}
Existing data placement schemes~\cite{maruf2023tpp,lee2023memtis,xiang2024nomad,vuppalapati2024tiered,zhou2024neomemhardwaresoftwarecodesigncxlnative} for dynamic data placement (i.e., migrating data based on workload demands) on {\chm} manage data at page granularity (e.g., 4\,KiB), such as MEMTIS~\cite{lee2023memtis}.
However, since tree nodes are typically smaller than pages, this granularity mismatch can lead to inaccurate hotness detection.
As shown in \autoref{fig:hotness_distribution}, even for the top 1\% of frequently accessed pages, the majority of nodes remain cold---the upper quartile of nodes' access count is just 16 and 4 for Masstree and ART, respectively.
Additionally, the hottest node on each page contributes 82\% and 85\% of total accesses, respectively.
Therefore, page-level migration may misplace cold nodes in fast memory, causing suboptimal data placement and increasing migration overhead.

\section{Design Principles and Challenges}
\label{sec:insight}

Our key insight is that the efficient placement of tree-structured indexes on {\chm} should match the tree's inherent characteristics with the features of {\chm}, which involves:
(1) matching the tree's node-granularity data usage with {\chm}'s cache-coherent memory semantics to achieve node-grained flexible data placement;
(2) matching node hotness indicated by the tree's hierarchical structure and access pattern with the performance gap between fast and slow memory to efficiently utilize fast and slow memory.

In this section, we present the principles of data placement derived from tree characteristics and the challenges in designing an efficient node-grained data placement scheme.

\subsection{Principles from Tree Characteristics}%

\subsubsection{Structural Characteristics.}%
Due to the tree's branching structure and the downward access from its root, nodes at upper levels (i.e., upper nodes) are fewer but have higher access frequency than lower nodes.
For instance, in a B+ tree~\cite{bplustree_code} and radix tree~\cite{art_code} with 50 million keys, nodes in the upper half levels comprise less than 1\% of total nodes but have average access frequency three orders of magnitude higher than those in the lower half levels.
Thus, \emph{upper nodes are naturally hot and well-suited for fast memory.}

\begin{figure}[t]
    \centering
    \subfloat[B+ tree]{
        \centering
        \includegraphics[width=0.49\linewidth]{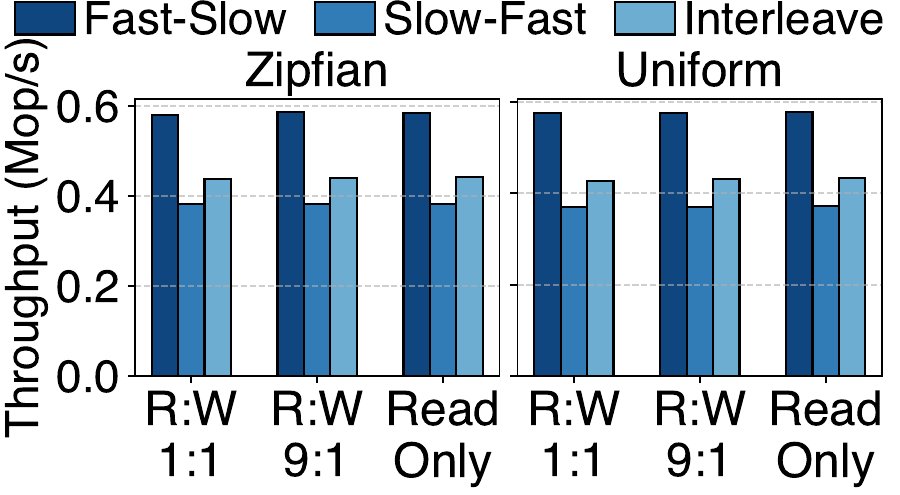}
        \vspace{-7pt}
        \label{fig:structural-btree}
    }
    \subfloat[Radix tree]{
        \centering
        \includegraphics[width=0.49\linewidth]{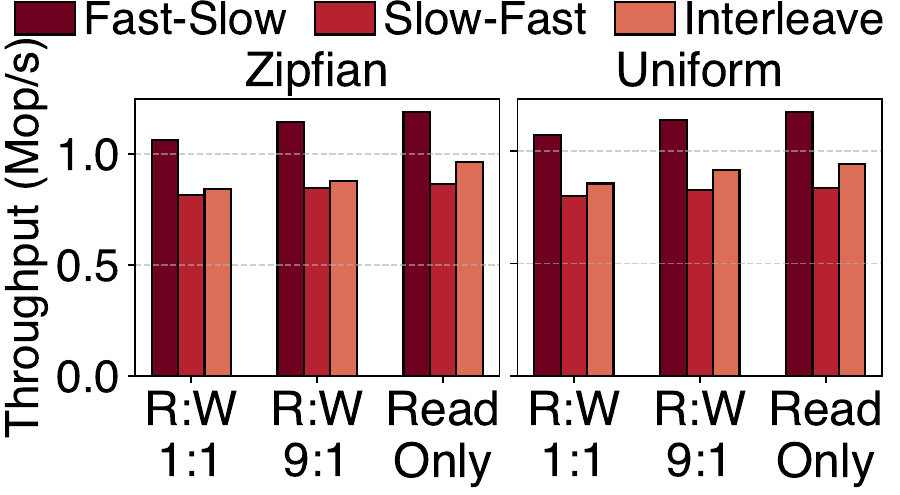}
        \vspace{-7pt}
        \label{fig:structural-art}
    }
    \vspace{-10pt}
    \caption{\textbf{Tput. of different node placement strategies.} 
    {\emph{The fast memory usage ratio is \textasciitilde50\% in all configurations.}}
    }
    \label{fig:structural-performance}
    \vspace{-14pt}
\end{figure}

We validate this observation using the B+ tree and radix tree in three settings: \emph{Fast-Slow} (upper nodes in fast memory), \emph{Slow-Fast} (the reverse of \emph{Fast-Slow}), and \emph{Interleave} (levels interleaved in two memory types).
\autoref{fig:structural-performance} shows \emph{Fast-Slow} performs best, as fewer upper nodes allow more nodes along the path to be placed in fast memory.
Notably, \emph{Interleave} outperforms \emph{Slow-Fast} but still trails \emph{Fast-Slow}, highlighting the benefits of placing upper nodes in fast memory.

\noindent
\textbf{Layer Principle.}
\label{insight:layer_insight}
It is beneficial to store the tree's upper nodes in the fast memory as many as possible.

\begin{figure}
    \centering
    \includegraphics[width=0.7\linewidth]{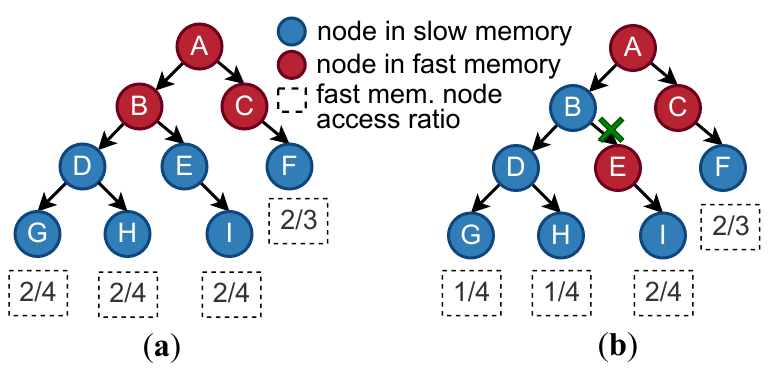}
    \vspace{-10pt}
    \caption{\textbf{\emph{Single-boundary structure}.} \emph{(a) The tree follows this structure. (b) When tree violates this, putting E in fast memory and B in slow memory, the fast memory access ratio of the paths from A to G and A to H decreases from $\frac{2}{4}$ to $\frac{1}{4}$.}}
    \label{fig:single_boundary}
    \vspace{-12pt}    
\end{figure}

Building on the \emph{layer principle}, we propose a guarantee to improve the expected ratio of fast memory access in each path to target nodes: \emph{all ancestors of a fast memory node should be placed in fast memory}.
As shown in \autoref{fig:single_boundary}(a), this guarantee ensures only one boundary between fast and slow memory nodes along all paths, so we term it as \emph{single-boundary structure}.
\autoref{fig:single_boundary}(b) illustrates that when the placement of nodes B and E violates this guarantee, the paths from A to G and H lose a fast memory node, potentially degrading the performance when accessing the data of nodes G and H.

\subsubsection{Access Characteristics.}%

Trees are accessed from the root downwards, forming access paths from the root to target nodes (e.g., leaf nodes in B+ trees and radix trees).
In real-world workloads, where accesses are often skewed~\cite{atikoglu2012workload,debrabant2013anti,eldawy2014trekking,tang2020xindex,yang2021large}, \emph{certain paths are accessed much more frequently.}
We identify these paths as \emph{hot paths}.

\begin{figure}[t]
    \centering
    \subfloat[B+ tree]{
        \centering
        \includegraphics[width=0.48\linewidth]{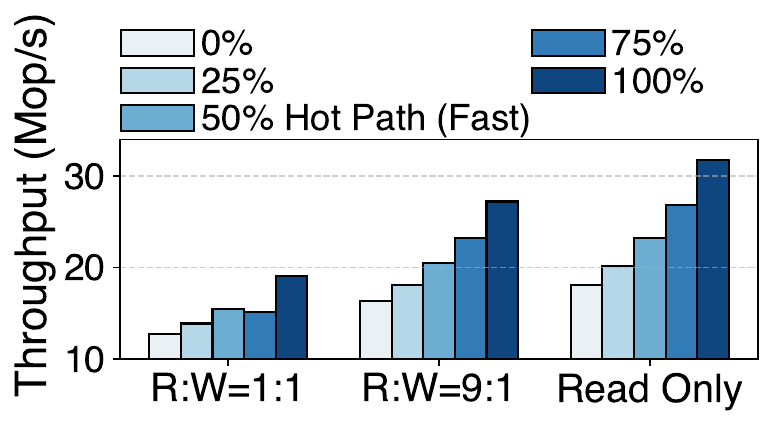}
        \vspace{-8pt}
        \label{fig:access-btree_layout}
    }
    \subfloat[Radix tree]{
        \centering
        \includegraphics[width=0.48\linewidth]{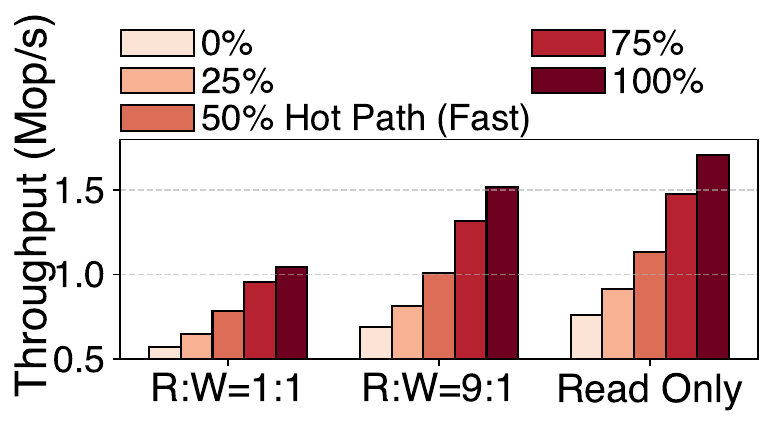}
        \vspace{-8pt}
        \label{fig:access-art_layout}
    }
     \vspace{-10pt}
    \caption{\textbf{Tput. at different hot path ratios in fast memory.} \emph{Percentages of \codeword{Hot Path (Fast)} indicate the proportion of hot paths in fast memory. For nodes not on \codeword{Hot Path (Fast)}, their placement follows the \codeword{Fast-Slow} config. in \autoref{fig:structural-performance}.}}
    \label{fig:hot_path_performance}
    \vspace{-12pt}
\end{figure}

To investigate how hot paths can affect the performance of trees on {\chm}, we evaluate a multi-threaded B+ tree~\cite{masstree_code} and the radix tree under a skewed workload, where 90\% of requests targeted 10\% data.
We vary the proportion of hot paths in fast memory and measure throughput.
As illustrated in \autoref{fig:hot_path_performance}, tree performance improves across all workloads with the higher proportion of hot paths in fast memory.

\noindent
\textbf{Path Principle.}
\label{insight:path_insight}
It is crucial to identify the hot paths and endeavor to place them in fast memory as many as possible.

\subsection{Challenges}%
\label{sec:challenge}

To adapt to dynamic workloads, an efficient data placement scheme should be able to (1) track node hotness and migrate nodes based on their hotness; and (2) adjust the allocation and migration strategy according to current memory usage.
Therefore, although these principles offer valuable guidance, several challenges still need to be addressed.

\noindent
\textbf{Fine-grained access tracking incurs high overhead for tree operations.}
Tree-structure indexes have high requirements for both latency and throughput~\cite{kocberber2013meet,zhang2016reducing,binna2018hot,zeitak2021cuckoo}.
Per-node access tracking introduces high overhead (over a 50\% performance drop) on the critical path.
This is because the tree's access pattern requires updating the access information of multiple nodes along the path for each data access.
Thus, we should relax the tracking granularity from nodes to paths.

\noindent
\textbf{Migration cannot guarantee single-boundary structure and may break the locality of hot paths.}
When using the path-grained tracking, existing migration mechanisms~\cite{lee2023memtis,raybuck2021hemem,maruf2023tpp,duraisamy2023towards,xiang2024nomad,vuppalapati2024tiered} that treat each object (e.g, page, node) independently are no longer suitable due to missing per-node hotness data.
Moreover, simply migrating by path is insufficient.
Since an internal node belongs to multiple paths, demoting it (i.e., moving to slow memory) may 
(1) break the \emph{single-boundary structure} if it has fast memory children; 
(2) disrupt hot path locality, as it may be part of other hot paths.

\noindent
\textbf{Allocation and migration require complicated control to meet the complexity of tree structure.}
Given the complexity of trees' hierarchical structure, more parameters are needed to control allocation and migration than page-level data placement schemes.
For example, migration requires two metrics for deciding the path and the number of nodes in each path that need to be migrated.
Additionally, allocation needs a threshold to determine whether a node should be placed in fast memory based on the \emph{layer principle}.
For these parameters, using static values cannot adapt to the dynamic workloads, and manual adjustments are error-prone and labor-intensive.
Moreover, compared to existing allocation and migration strategies that are either passive (e.g., demotion only when fast memory is scarce~\cite{lee2023memtis,xiang2024nomad}) or lack coordination between allocation and migration~\cite{maruf2023tpp}, we need proactive and holistic strategies to meet the performance stability requirements of the trees.

\section{Design of {\sys}}%
\label{sec:detail}

\begin{figure}[t]
    \centering
    \includegraphics[width=0.85\linewidth]{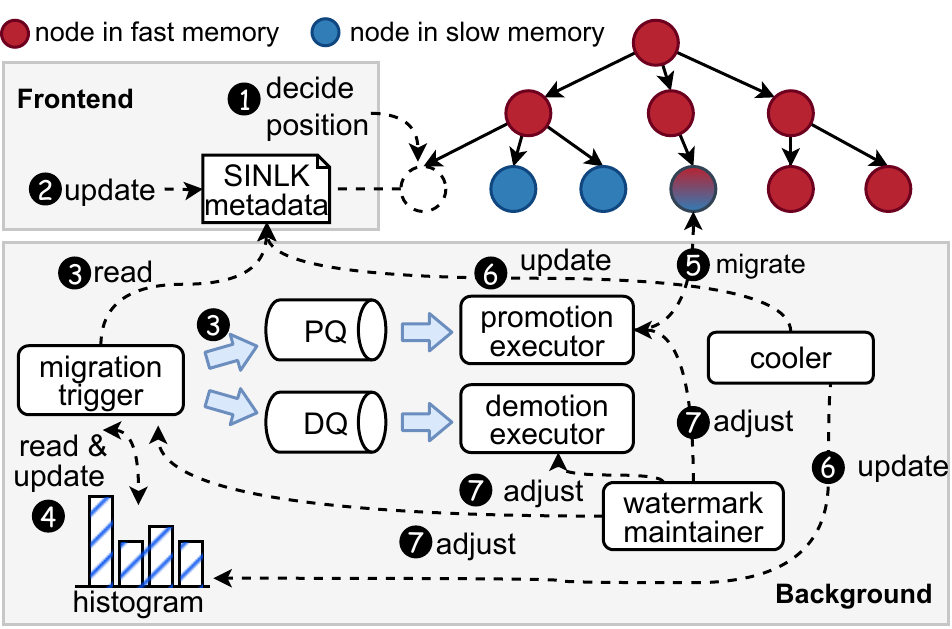}
    \vspace{-10pt}
    \caption{\textbf{Architecture and interactions of {\sys}.}{\emph{PQ stands for promotion queue, DQ stands for demotion queue.}}}%
    \label{fig:high-level-arch}
    \vspace{-15pt}
\end{figure}

To address the challenges in \autoref{sec:challenge}, we design {\sys}, an efficient and generic data placement scheme for tree-structure indexes on {\chm}.
As depicted in \autoref{fig:high-level-arch}, {\sys} consists of a lightweight frontend module and a background module.

\noindent
\textbf{The frontend module} is invoked in the critical path of the tree operations so that it is designed to be lightweight to minimize its impact on performance.
It determines the initial position of a new node (\ding{182}) based on the \emph{layer principle} while maintaining the \emph{single-boundary structure}.
It also updates the access frequency of leaf nodes (\ding{183}), laying the groundwork for hot path and cold node identification.

\noindent
\textbf{The background module} consists of several background workers to conduct node migration (\autoref{sub:migration}) and system adjustment (\autoref{sub:watermark}) asynchronously to optimize fast memory usage.
The \emph{migration trigger} decides whether migration is needed and picks the appropriate nodes for migration based on their access frequency (\ding{184}) and the distribution (i.e., the histogram) of all leaf nodes' access frequency (\ding{185}).
The \emph{promotion executor} and \emph{demotion executor} get nodes from the corresponding queues and carry out the migration tasks (\ding{186}).
The \emph{cooler} periodically halves the nodes' access frequency (\ding{187}) to ensure the freshness of access information.
To coordinate all workers, the \emph{watermark maintainer} monitors the fast memory usage and adjusts system parameters and the execution of other workers to ensure efficient utilization of fast memory (\ding{188}).

To reduce the impact of the data placement scheme on tree performance (Challenge 1), we propose a lightweight node allocation and tracking scheme (\autoref{sub:frontend}). 
To maintain the \emph{single-boundary structure} and localize hot paths during node migration (Challenge 2), we introduce the structure-aware migration procedures (\autoref{sub:migration}). 
Additionally, we holistically adjust allocation and migration using the hyper watermark mechanism (\autoref{sub:watermark}) to ensure stable high performance while optimizing fast memory utilization (Challenge 3).

\subsection{Lightweight Node Allocation and Tracking}
\label{sub:frontend}
{\sys} only leaves two essential and lightweight tasks on the critical path of tree operations: 
Determining each node's initial position based on the \emph{layer principle} while maintaining the \emph{single-boundary structure}, and tracking access frequency for each path through leaf nodes as guided by the \emph{path principle}.
All other tasks are executed in the background asynchronously, minimizing the impact on trees' performance.

\noindent\textbf{Layer-aware Allocation.}
Guided by the \emph{layer principle}, {\sys} allocates upper nodes in fast memory while ensuring that the relationship between the new node and its ancestors conforms to the \emph{single-boundary structure}.
Specifically, {\sys} maintains the highest level of fast memory node allocation, $L_{fast}$, which is adjusted based on current fast memory usage.
When allocating new nodes or changing the node's position by auto-balancing, the decision on the node's placement is based on its current level $l$ and its parent's type.
If $l < L_{fast}$ and its parent is in fast memory, the node is allocated in fast memory; otherwise, it is placed in slow memory. 
Such a layer-aware allocation incurs little overhead on the critical path and effectively maintains the \emph{single-boundary structure}.

\noindent\textbf{Leaf-centric Access Tracking.}
The \emph{path principle} highlights the necessity of placing hot paths in fast memory.
Since data access ultimately reaches the leaf node, we can identify the path leading to the frequently accessed leaf as the hot path.
Essentially, determining hot paths is based on leaf nodes' access frequency.  

Therefore, {\sys} maintains access frequency information only for leaf nodes.
For each data access, instead of updating the access information of all nodes along the path, {\sys} only updates the access frequency information of the leaf node.
This leaf-centric approach significantly reduces the overhead of access tracking while providing sufficient information to identify hot paths and cold nodes for effective migration.

\subsection{Structure-Aware Migration}
\label{sub:migration}

{\sys} designs the migration mechanism tailored for tree structures, ensuring that the hot paths can be quickly migrated to fast memory (\autoref{sub:promotion}) and that node demotion does not disrupt the \emph{single-boundary} structure (\autoref{sub:demotion}). 
Additionally, {\sys} employs a histogram-based approach to precisely identify hot paths and cold nodes (\autoref{sub:histogram}).
To achieve efficient migration, {\sys} decouples the preparation and execution of migration processes, assigning the migration trigger and promotion/demotion executors to work collaboratively (\autoref{sub:migration_workers}).

\begin{figure}[t]
    \centering
    \includegraphics[width=0.82\linewidth]{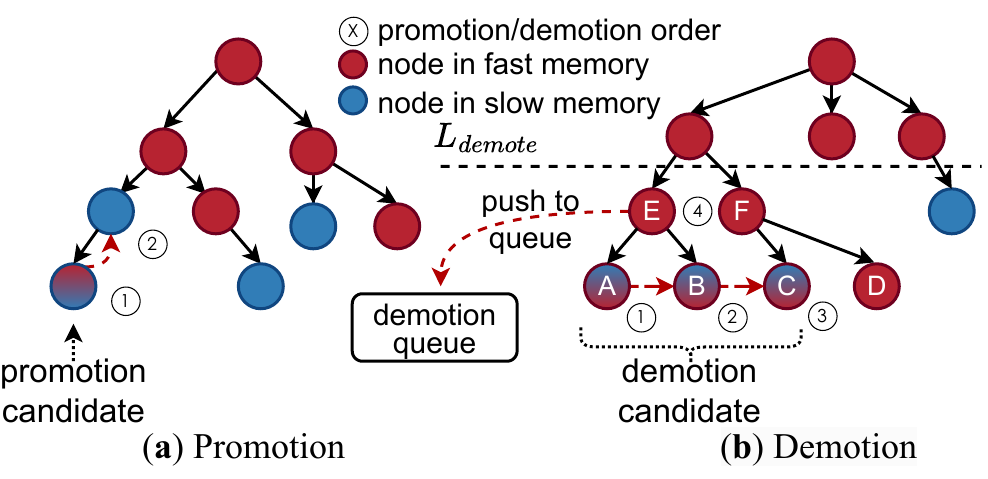}
    \vspace{-12pt}
    \caption{\textbf{Promotion and demotion procedures.}}%
    \label{fig:migration_mechanism}
    \vspace{-18pt}
\end{figure}

\subsubsection{Promotion Procedure.}
\label{sub:promotion}

When a leaf node in slow memory is identified as a hot node that needs promotion, following the \emph{path principle}, it is necessary to quickly migrate the entire hot path to fast memory.
As shown in \autoref{fig:migration_mechanism}(a), {\sys}'s promotion procedure starts with the leaf node (\ding{192}), recursively migrating all of its ancestor nodes in slow memory (\ding{193}) upwards to fast memory.

\subsubsection{Demotion Procedure.}
\label{sub:demotion}

When a leaf node in fast memory is identified as a cold node, it is essential to demote not only this leaf node but also some of its ancestors to better utilize the limited fast memory resources.
The specific number of ancestor nodes to be migrated depends on two key factors.
The first is the system parameter $L_{demote}$, which reflects the intensity of demotion required under the current fast memory usage.
The second factor is to ensure that the demotion does not disrupt the \emph{single-boundary structure}.
As illustrated in \autoref{fig:migration_mechanism}(b), nodes A, B, and C are current demotion candidates. 
Migrating node C and then its parent node F breaches the \emph{single-boundary structure} because node D is still in fast memory.
Therefore, to demote parent node F, node D must also be demoted simultaneously.
However, if node D, which is not a demotion candidate at that time, is demoted, it may inadvertently migrate a node in a hot path to slow memory, potentially degrading system performance.

To address this, {\sys} adopts the demotion procedure as depicted in \autoref{algo:demotion}: once a node is demoted, its parent node is queued for demotion (\autoref{line:parent}).
The decision to demote the parent node is deferred until all its cold children have been processed by the demotion executor.
If all of the parent node's children have been in slow memory by then, it indicates that the parent node is not part of any hot path, making it safe to demote.
Conversely, if not all children are in slow memory, the parent node is kept in fast memory to prevent the demotion of hot paths while maintaining the \emph{single-boundary structure} (\autoref{line:continue}).
Additionally, the lowest level for demotion is denoted by $L_{demote}$, nodes below $L_{demote}$ will not participate in the demotion process (\autoref{line:if-start}).
Therefore, for the tree in \autoref{fig:migration_mechanism}(b), the demotion process follows the labels \ding{192}\textasciitilde\ding{195}.

\setlength{\textfloatsep}{10pt plus 1.0pt minus 2.0pt}
\begin{algorithm}[t]
    \setlength{\belowdisplayskip}{3pt}
    \caption{Demotion Algorithm}
    \label{algo:demotion}
    \small
    \SetAlgoLined
    \textbf{Initialization:} The migration trigger traverses all leaf nodes, finds the leaf nodes that need to be demoted, and adds them to the demotion queue\;
    \While{demotion queue is not empty}{
        pop node $cur$ from demotion queue\;
        \nl\If{$cur$.level < $L_{demote}$}{\label{line:if-start}break\;}
        \nl\If{$cur$ is internal node \textbf{and} at least one of its children is in fast memory} {\label{line:continue}{continue\;}}  
        demote $cur$ to slow memory\; 
        add $cur$'s parent to the demotion queue if it is not in the queue\; \label{line:parent}
    }
\end{algorithm}

\subsubsection{Histogram-based Hotness Identification.}
\label{sub:histogram}

Like prior studies~\cite{lagar2019software,lee2023memtis,xu2024flexmem}, {\sys} employs a logarithmic histogram to represent access frequency distribution, thereby identifying hot paths and cold nodes for migration.
 
The migration trigger and the cooler collaborate to maintain the histogram.
The migration trigger updates the histogram with leaf nodes' latest access frequency.
The cooler halves all leaf nodes' access frequency and reflects the reduction in the histogram.
Note that as the histogram is on a logarithmic scale, the reduction can be efficiently conducted by shifting all bins to the left by one position.

Based on the histogram, {\sys} uses a hot threshold $T_{hot}$ and a cold threshold $T_{cold}$ to identify hot paths and cold nodes for migration.
Leaf nodes with access frequency higher than $T_{hot}$ are leaves of hot paths; while leaf nodes with access frequency lower than $T_{cold}$ are deemed cold nodes.

As static thresholds cannot adapt to dynamic workloads, {\sys} periodically updates the hot and cold thresholds using two parameters, $P_{cold}$ and $P_{hot}$.
These two parameters\footnote{$P_{cold}$ and $P_{hot}$ are initialized based on the maximum fast memory usage and further dynamically adjust through the hyper watermark mechanism (\autoref{sub:watermark}) based on the current fast memory usage.} determine the percentage of cold and hot leaves, respectively.
{\sys} scans the histogram to set $T_{hot}$ to a value where nodes with frequency higher than $T_{hot}$ just exceeds $P_{hot}$.
Similarly, $T_{cold}$ is set to a value where nodes with frequency lower than $T_{cold}$ just fall below $P_{cold}$.
Note that the interval between $T_{hot}$ and $T_{cold}$ prevents premature classification of less accessed nodes as cold, avoiding aggressive node migration.

\begin{figure}[t]
    \centering
    \includegraphics[width=0.85\linewidth]{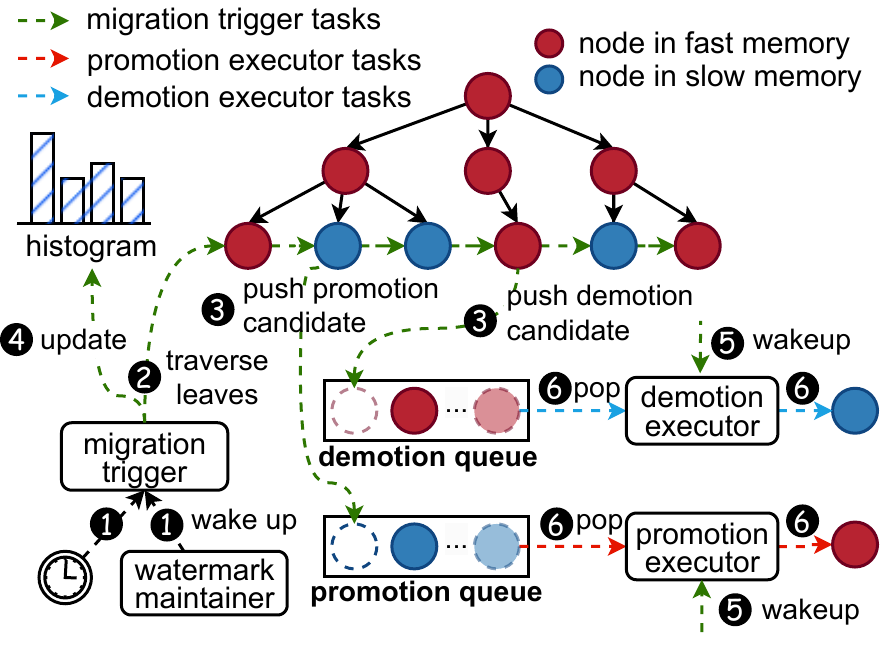}
    \vspace{-15pt}
    \caption{\textbf{Migration workflow.}}%
    \label{fig:migration}
    \vspace{-4pt}
\end{figure}

\subsubsection{Migration Workflow.} 
\label{sub:migration_workers}

As shown in \autoref{fig:migration}, the migration trigger is activated when fast memory usage is under pressure or on a schedule (\ding{182}).
It scans all leaf nodes (\ding{183}) and selects cold ones in fast memory as demotion candidates, and hot ones in slow memory as promotion candidates.
These nodes are pushed into respective demotion/promotion queue (\ding{184}). 
During scanning, the migration trigger also updates the histogram recording access frequency distribution (\ding{185}).
When the queue is too long or the migration trigger has checked all leaf nodes, the corresponding executor is awakened (\ding{186}) to carry out the migration according to the procedures described in \autoref{sub:promotion} and \autoref{sub:demotion} (\ding{187}).
It is worth noting that migration is executed by independent workers, enabling the trigger to continue selecting migration candidates without being blocked by the migration process.

\subsection{Hyper Watermark Mechanism}
\label{sub:watermark}

To achieve high performance, {\sys} selects appropriate initial positions for nodes based on their level (\autoref{sub:frontend}) and periodically migrates cold nodes and hot paths (\autoref{sub:migration}).
These procedures involve a series of tree-related parameters, including $L_{fast}$ (\autoref{sub:frontend}), $P_{cold}$/$P_{hot}$ (\autoref{sub:migration}), and $L_{demote}$ (\autoref{sub:migration}).
Manually adjusting these parameters is challenging, as it is difficult to adapt to workload changes and may cause conflicting fast memory usage intentions between node migration and allocation, causing performance instability.

Therefore, {\sys} introduces a hyper watermark mechanism that synthetically adjusts these parameters and coordinates the behavior of background workers based on the current fast memory usage.
It ensures consistent fast memory usage trends for node allocation and migration under various conditions. 
By trying to maximize the utilization of fast memory without exhausting it, the hyper watermark mechanism effectively prevents performance degradation from burst migrations and other extreme behaviors.

\subsubsection{Asymmetric Adjustment.}
\label{sub:asym-adjustment}

The impacts of fast memory strain and abundance are different.
Depletion of fast memory can force the new upper-level nodes to be allocated to slow memory, obstruct hot path promotions, or trigger burst demotions, resulting in severe performance degradation that is unacceptable for high-performance tree-structure indexes.
Conversely, when fast memory is abundant, although it leaves some fast memory underutilized, it does not have such a significant impact on performance stability and is advantageous for handling burst insertions of new nodes and promoting hot paths in the future.

Therefore, {\sys} employs asymmetric adjustments.
When the fast memory usage ratio exceeds the high watermark $U_{high}$, {\sys} adopts an aggressive adjustment strategy to promptly bring the fast memory usage back to a reasonable range.
Conversely, when the fast memory usage ratio falls below the low watermark $U_{low}$, a conservative adjustment is taken to optimize fast memory usage while reserving space for new nodes and hot path promotions.

\begin{figure}[t]
    \centering
    \includegraphics[width=0.95\linewidth]{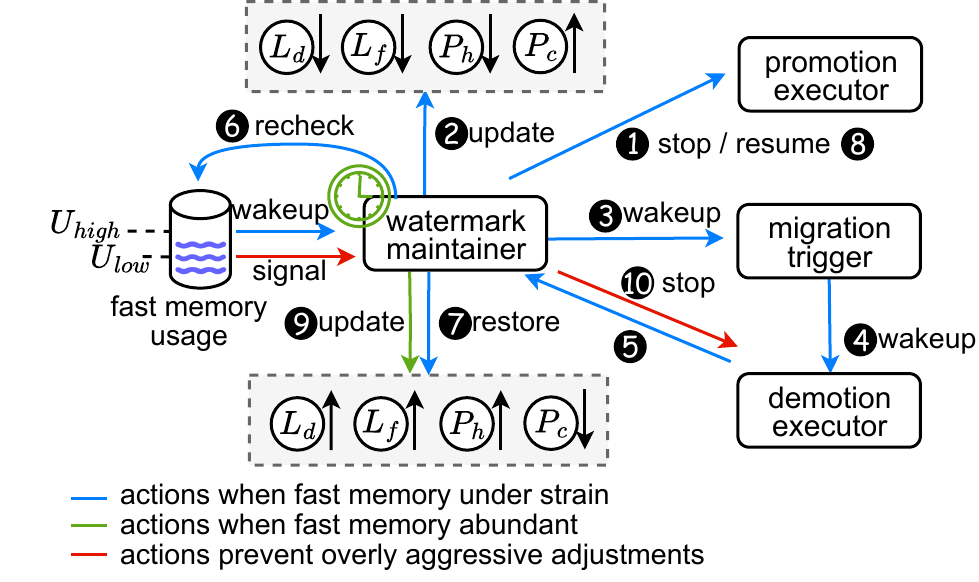}
    \vspace{-12pt}
    \caption{\textbf{System parameter adjustments and the collaboration of workers in the hyper watermark mechanism.} \emph{$L_{d}$ is $L_{demote}$, $L_{f}$ is $L_{fast}$, $P_{c}$ is $P_{cold}$, and $P_{h}$ is $P_{hot}$.}%
    }%
    \label{fig:watermark}
    \vspace{-5pt}
\end{figure}

\subsubsection{Fast Memory Under Strain.}

When the fast memory usage ratio exceeds $U_{high}$, the watermark maintainer is immediately awakened to regulate the system parameters and background workers.
The blue lines in \autoref{fig:watermark} illustrate this process.
First, the watermark maintainer temporarily halts all ongoing promotion tasks (\ding{182}).
Subsequently, it adjusts system parameters to increase demotion likelihood by increasing $P_{cold}$ and decreasing $P_{hot}$, intensify demotion by decreasing $L_{demote}$, and reduce the number of nodes initialized in fast memory by decreasing $L_{fast}$ (\ding{183}).
The watermark maintainer then awakens the migration trigger to select cold nodes for demotion (\ding{184}), which in turn activates the demotion executor to migrate these cold nodes (\ding{185}).
Following this, the watermark maintainer waits for the demotion executor to complete the demotion tasks (\ding{186}), then rechecks the fast memory usage (\ding{187}), and repeats steps \ding{183}--\ding{187} until the fast memory usage ratio falls below $U_{high}$.
Upon achieving this, the watermark maintainer resets these system parameters to their pre-adjustment states (\ding{188}) and re-enables the execution of the promotion executor (\ding{189}).

\subsubsection{Fast Memory Abundant.}

As depicted by the green line and clock icon of \autoref{fig:watermark}, the watermark maintainer periodically checks if the fast memory usage ratio drops below $U_{low}$.
If so, it increases $P_{hot}$, $L_{demote}$, $L_{fast}$; and decreases $P_{cold}$ (\ding{190}).
This raises chances of promotion and fast memory allocation while reducing demotion likelihood.
This approach is deemed conservative as it only adjusts parameters without instantly triggering migration executions.

\subsubsection{Optimizations.}

When fast memory is under strain, {\sys} uses an aggressive adjustment strategy that sometimes over-identifies cold nodes, leading to excessive demotions and potentially dropping the fast memory usage ratio below $U_{low}$.
To mitigate this and protect performance, {\sys} acts promptly when the fast memory usage ratio falls below $U_{low}$ and adjustments for prior memory pressure are still in progress.
The watermark maintainer is signaled to halt current demotion tasks (\ding{191}).
Subsequently, {\sys} restores system parameters to their original states (\ding{188}) and reactivates the promotion executor (\ding{189}), safeguarding system performance through efficient fast memory utilization.

When fast memory is under pressure, the watermark maintainer may need multiple attempts, involving several adjustments to system parameters.
If the adjustments to $L_{demote}$ and $L_{fast}$ are too aggressive, they can lead to excessive demotion and an overly pessimistic fast memory usage trend, hurting performance.
To prevent this, we limit changes to $L_{demote}$ and $L_{fast}$ based on the maximum fast memory usage, avoiding excessively aggressive adjustments.

The demotion procedure in \autoref{sub:demotion} only considers demoting the ancestors of leaf nodes in fast memory. 
To prevent cold ancestors of slow memory leaf nodes from being stuck in fast memory and to demote more cold internal nodes under fast memory strain, the migration trigger adds some cold leaf nodes in slow memory to the demotion queue, creating the opportunity to demote their ancestors later.

\section{Implementation}%
\label{sec:impl}

\noindent
\textbf{{\sys} Framework.}
We provide a lightweight and easy-to-use framework for integrating {\sys} into various existing tree-structure indexes.
For {\sys} metadata, the framework requires developers to add one byte in internal nodes to record the node's level and memory type.
Leaf nodes need two additional bytes for the access frequency.
This memory overhead is acceptable because tree nodes are typically equal to or larger than the cacheline~\cite{hankins2003effect}, and sometimes the node size is unaffected due to memory alignment.
For the frontend module, {\sys} framework provides helper functions that implement layer-aware allocation and leaf-centric access tracking for developers to use.
Additionally, developers need to implement tree-specific logic in the interfaces defined by {\sys} framework for the background module to interact with the tree.
Notably, while {\sys} requires some user-implemented functions, many of them can leverage the existing logic of the tree, such as node splitting for migration and the iterator for leaf traversal.

\noindent
\textbf{Integration.}
We have integrated ART and Masstree (a B+ tree variant optimized for multicore architecture) with {\sys} framework, termed as {\sart} and {\smass}, respectively.
In contrast to Masstree's 25\,k LoC and ART's 1.7\,k LoC, {\sys} only modifies  \textasciitilde150 and \textasciitilde50 LOC respectively.
Moreover, we use \textasciitilde1.8\,k LoC to implement the framework, along with \textasciitilde450 and \textasciitilde320 LoC to implement its interfaces for Masstree and ART, respectively.
This lightweight integration shows that {\sys} can be easily applied to existing tree-structure indexes.

\noindent
\textbf{Concurrency Control.}
To prevent concurrent writes to the migrating node, we lock the node and its parent until the node's migration completes.
For concurrent reads, we use an unused version bit to indicate the node's migration status.
Read correctness is ensured by the optimistic read, which checks version numbers before/after each read and retries if changed.
Note that there is no lock contention between background workers, which use atomic operations (e.g., CAS) to access shared state (e.g., parameters, histogram).

\noindent
\textbf{Configuration.}
In {\sys}'s runtime, the migration trigger, the cooler, and the watermark maintainer wake up every 500\,ms, 2000\,ms, and 100\,ms, respectively. The high watermark and low watermark are set to 95\% and 85\%, respectively.

\section{Evaluation}%
\label{sec:eval}

\begin{table}[t]
    \centering
    \caption{\textbf{Workload composition.} {\emph{SP is Skewed Partition.}}}
    \vspace{-10pt}
    \resizebox{\linewidth}{!}{
        \begin{tabular}{cc}
            \toprule        
                        \textbf{Workload} & \textbf{Request Composition}   \\ \hline
            Update Heavy (SP, YCSB-A)              & 50\% Read, 50\% Update \\
            Read Mostly (SP, YCSB-B)               & 95\% Read, 5\% Update  \\
            Read Only (SP, YCSB-C)                 & 100\% Read             \\
            With Insert (SP), Read Latest (YCSB-D) & 95\% Read, 5\% Insert  \\
            Short Ranges (YCSB-E)                  & 95\% Scan, 5\% Insert  \\
            Read Modify Write (RMW) (YCSB-F)             & 50\% Read, 50\% RMW    \\
            \bottomrule
        \end{tabular}
    }
    \label{tab:workload}
    \vspace{-2pt}
\end{table}

\begin{figure*}[t]
    \centering
    \subfloat[Throughput]{
        \includegraphics[width=0.42\textwidth]{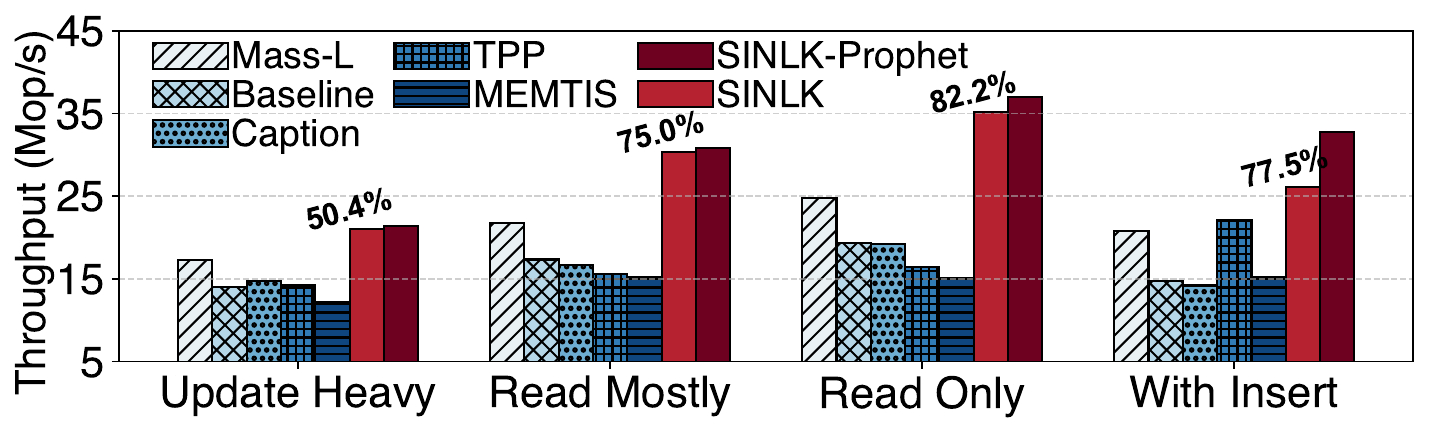}
        \vspace{-8.2pt}
        \label{fig:micro_throughput}}
    \subfloat[Latency of \emph{Update Heavy} workload]{
            \includegraphics[width=0.46\textwidth]{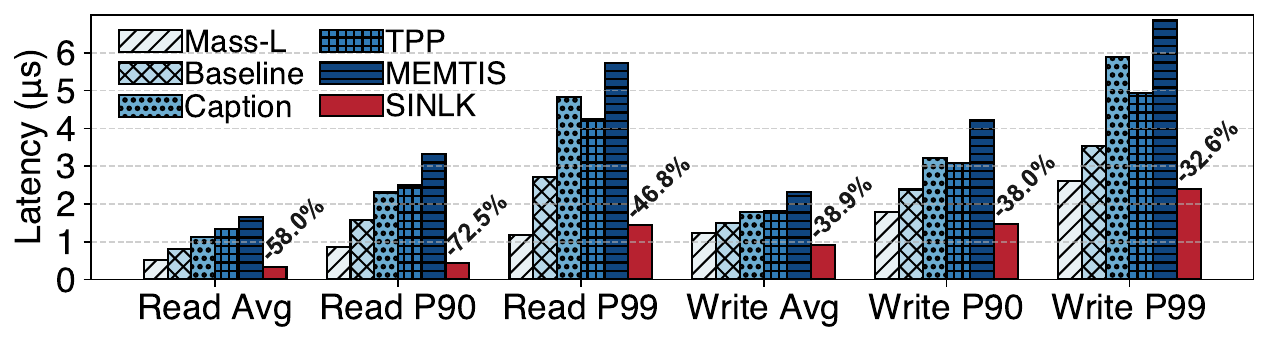}
            \vspace{-8pt}
            \label{fig:latency}}
    \vspace{-10pt}
    \caption{\textbf{Performance of {\sys} (S-Masstree) and compared systems in the microbenchmark.}}
    \label{fig:microbenchmark}
\end{figure*}

\begin{figure*}[t]
    \centering
    \subfloat[S-ART]{
        \includegraphics[width=0.48\linewidth]{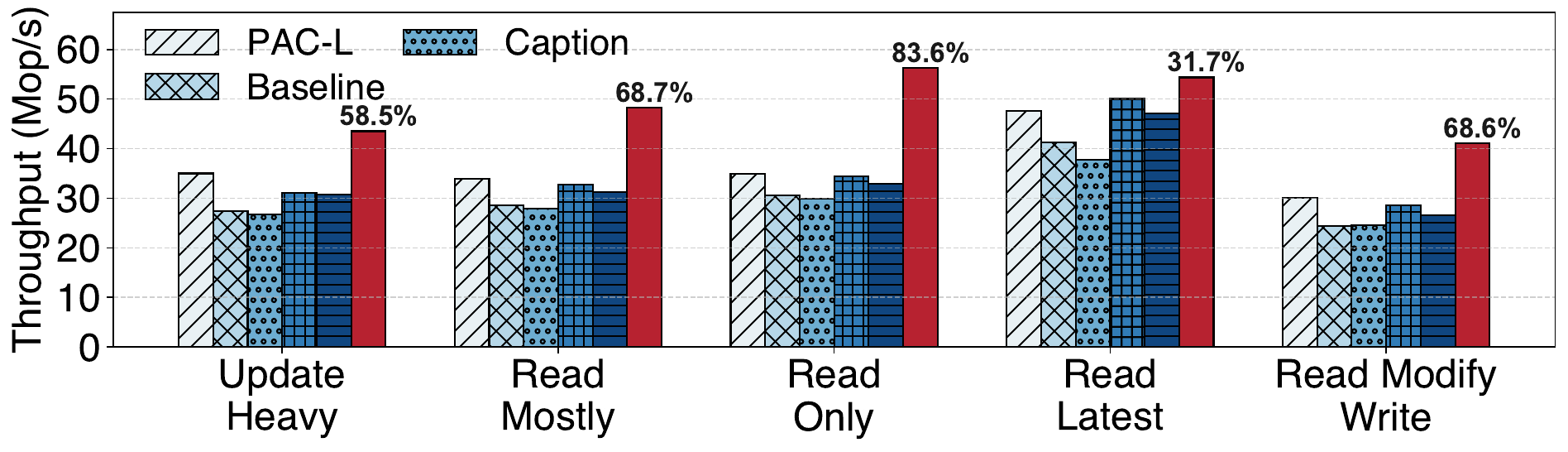}
        \vspace{-7pt}
        \label{fig:art_throughput}}
    \subfloat[S-Masstree]{
        \includegraphics[width=0.48\linewidth]{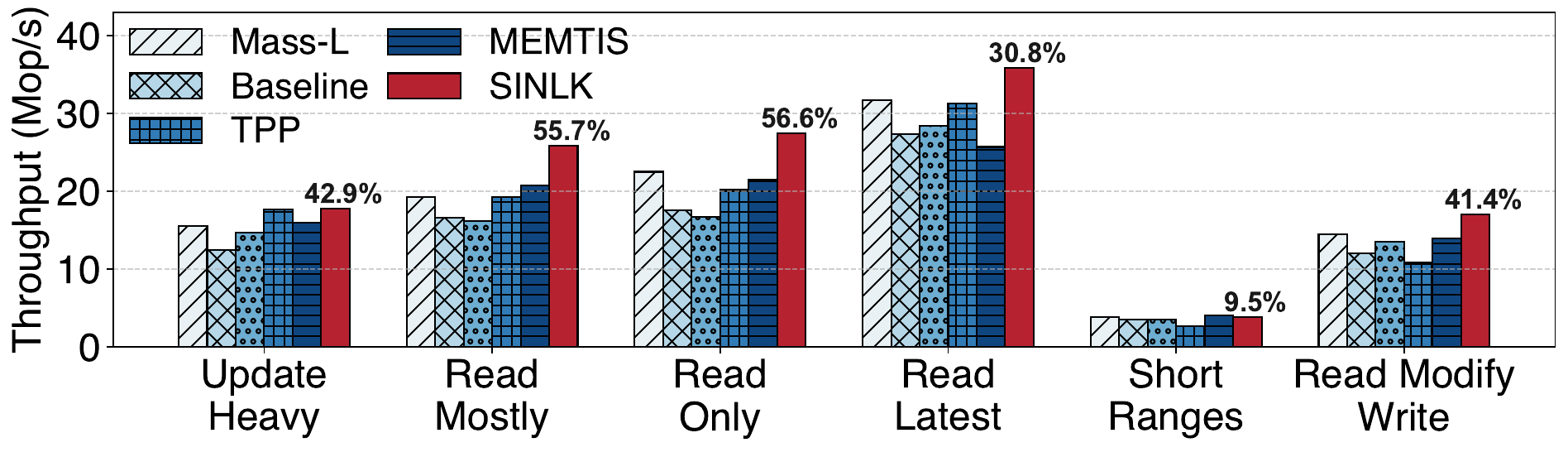}
        \vspace{-7pt}
        \label{fig:masstree_throughput}}
     \vspace{-10pt}
    \caption{\textbf{Throughput of {\sys} and compared systems in the macrobenchmark.}}
    \label{fig:macrobenchmark}
\end{figure*}

In this section, we evaluate {\sys} from multiple perspectives to answer the following questions:
\begin{itemize}[leftmargin=1em,topsep=-3pt]
\item Does {\sys} perform well in various scenarios? (\autoref{sec:overall-performance})
\item Can {\sys} maintain good performance in dynamic changing workloads and different initial system settings? (\autoref{sec:sensitivity-analysis})
\item How scalable is {\sys} when varying the number of threads and dataset sizes? (\autoref{sec:scalability-analysis})
\item How do {\sys}’s techniques contribute to the final performance, and how does {\sys} perform across various key/value sizes? (\autoref{sec:performance-breakdown})
\item How significant is {\sys}'s overhead? (\autoref{sec:overhead-analysis})
\item How does {\sys} perform in real-world workload? (\autoref{sec:real-world-eval})
\end{itemize}

\subsection{Environment Setup}
\label{sec:eval-setup}

\textbf{Testbed.} 
We run all experiments on Intel\textsuperscript{\textregistered} Xeon\textsuperscript{\textregistered} Platinum 8468V CPUs (48 cores), equipped with 395\,MiB CPU cache, 32\,GiB DRAM, and 32\,GiB CXL-attached DRAM (connected through a CXL memory expansion card~\cite{cxl_card}). 
We use a single socket to avoid NUMA effects. 
All experiments except the concurrency scalability are conducted with 28 threads.

\noindent
\textbf{Workloads.} 
We use the Skewed Partition (SP) synthetic workload as the microbenchmark, where 90\% of requests target 5\% of contiguous keys (termed \emph{hot region}).
This pattern closely matches the strong locality of hot keys in production environments~\cite{cao2020characterizing}.
We use YCSB\footnote{We don't evaluate \emph{Short Range} for S-ART as its source code lacks scan.}~\cite{YCSB2010} with the default Zipfian distribution as the macrobenchmark.
Details of both benchmarks are shown in \autoref{tab:workload}. 
Additionally, we evaluate {\sys} using Alibaba Block Traces~\cite{alibaba_block_traces,wang2022separating} which is a real-world workload collected from production.

\noindent
\textbf{Compared Systems.} 
The baseline uses the default weighted allocation policy~\cite{linux_weighted_interleave} to allocate a specified ratio of memory to CXL-attached DRAM. 
Moreover, we compare {\sys} with three SOTA data placement schemes for {\chm}: TPP~\cite{maruf2023tpp}, MEMTIS~\cite{lee2023memtis}, and Caption~\cite{sun2023demystifying}.
For {\sart}, we compare it with PAC-L, the optimized PACTree~\cite{kim2021pactree} with small leaves and fast memory internode (\autoref{sub:fast_memory_internode}).
For {\smass}, we compare it with Mass-L (i.e., Masstree with fast memory internode\footnote{Internodes are allocated to slow memory when fast memory runs out.}).
As Caption does not specify the maximum fast memory usage, we set its limit to be the same as the baseline.

\subsection{Overall Performance}
\label{sec:overall-performance}

\subsubsection{Microbenchmark.} 

{\sys}-Prophet represents an ideal scenario where all hot paths are allocated in fast memory during initialization. 
To ensure fair comparison, we measure {\sys}-Prophet's fast memory usage (\textasciitilde10\% of total memory) and use this fast memory amount as the limit for other systems.
\autoref{fig:micro_throughput} shows that S-Masstree outperforms the baseline by 50\%--82\%\footnote{ In \autoref{fig:microbenchmark}, \autoref{fig:macrobenchmark}, \autoref{fig:real_world_throughput_art} and \autoref{fig:real_world_throughput}, the marked improvements are compared with the baseline.}.
Furthermore, {\sys} performs similarly to {\sys}-Prophet in the first three workloads, showing its ability to effectively identify and promote hot paths at runtime.
For \emph{With Insert}, {\sys}-Prophet outperforms {\sys} because it allocates most new nodes in fast memory, exceeding {\sys}'s maximum fast memory usage by 1.3--1.5$\times$.

For other systems, Caption performs similarly to the baseline due to its interleave allocation method, which is similar to the baseline.
TPP performs well in \emph{With Insert} because it allocates all nodes in fast memory at the beginning when fast memory is abundant. 
However, TPP struggles to identify cold nodes and demote them in read-heavy workloads, hurting performance. 
MEMTIS performs comparatively poorly across all workloads because its default usage of huge pages does not adapt well to workload characteristics.
{\massl} outperforms other baselines but still lags behind {\sys} as static placement cannot adapt to dynamic workloads.

\begin{figure*}[t]
    \centering
    \subfloat[Dynamic workload]{
        \includegraphics[width=0.29\textwidth]{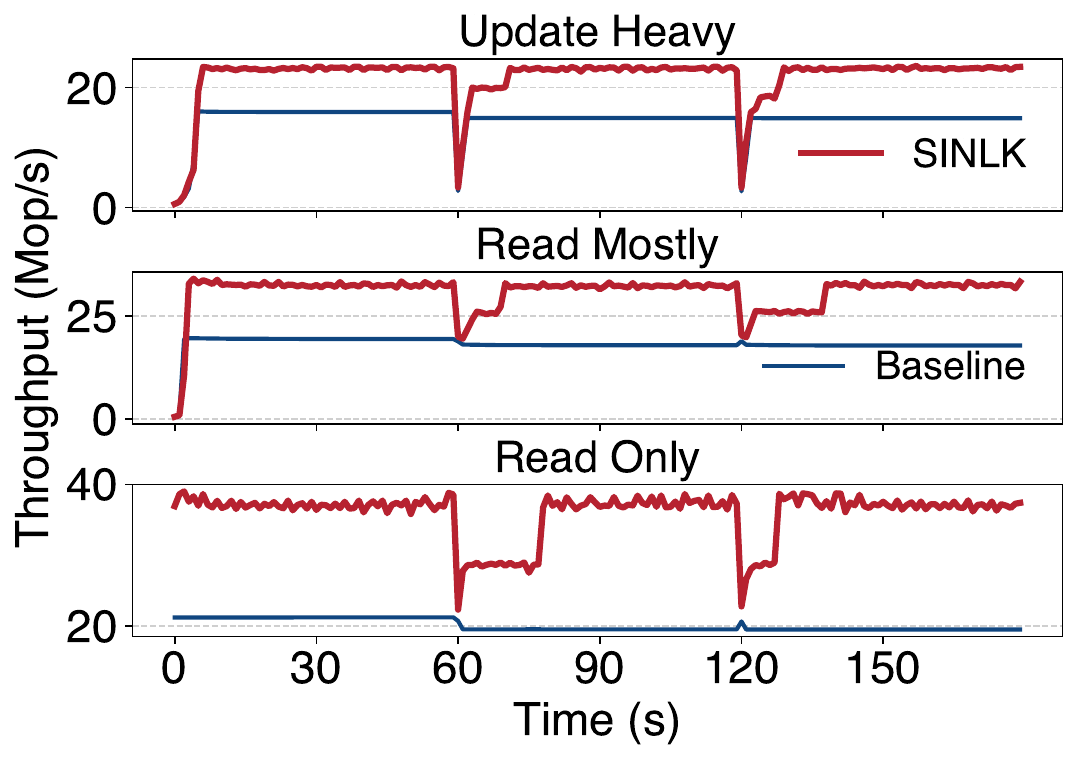}
        \vspace{-5pt}
        \label{fig:dynamic_hot_range}}
    \subfloat[Different worker wake-up intervals]{
        \includegraphics[width=0.32\textwidth]{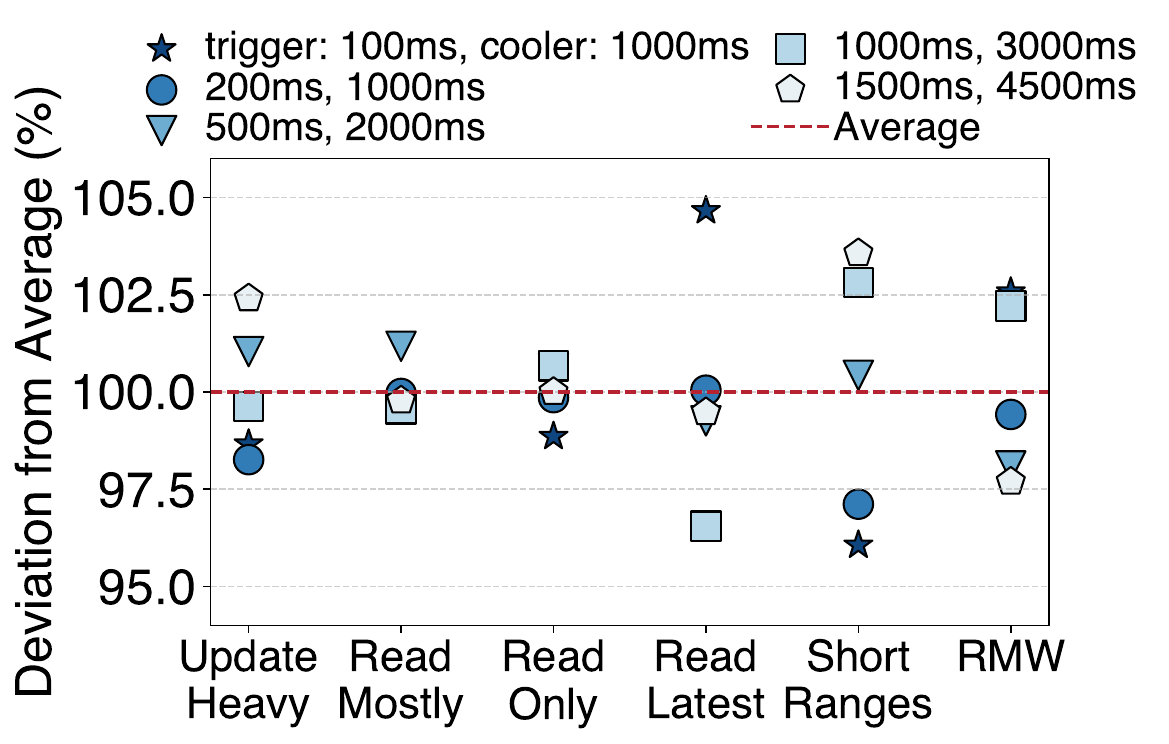}
        \vspace{-5pt}
        \label{fig:diff_wakeup_interval}}
    \subfloat[Different maximum fast memory usage]{
        \includegraphics[width=0.35\textwidth]{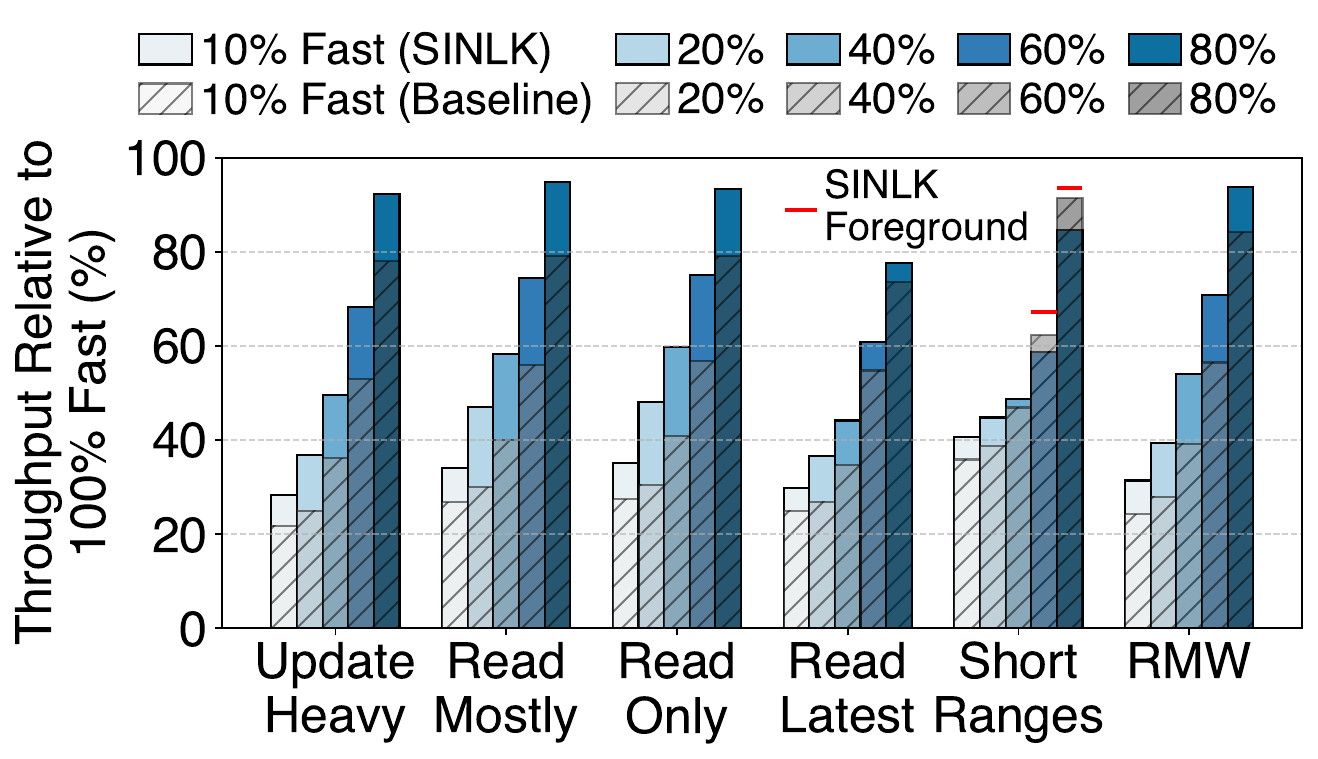}
        \vspace{-5pt}
        \label{fig:diff_memory_usage}}
     \vspace{-12pt}
    \caption{\textbf{Sensitivity analysis of {\sys} ({\smass}).}}
    \label{fig:sensitivity}
\end{figure*}

\begin{figure*}[t]
    \centering
    \subfloat[Scalability with different thread counts]{
        \includegraphics[width=0.62\linewidth]{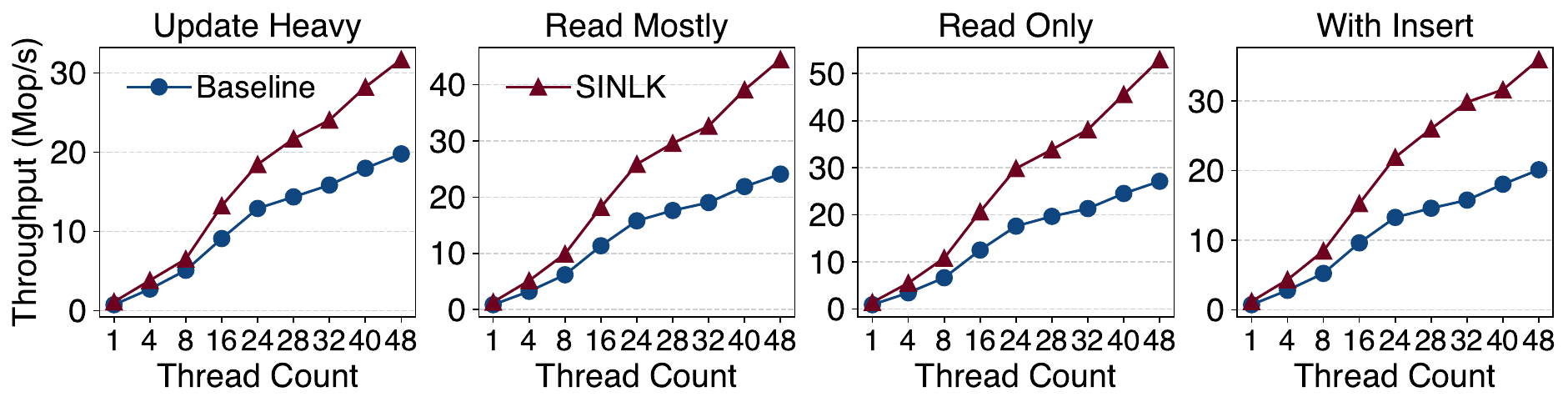}
        \vspace{-8pt}
        \label{fig:concurrency-scalability}}
    \subfloat[Scalability with different dataset sizes]{
        \includegraphics[width=0.365\linewidth]{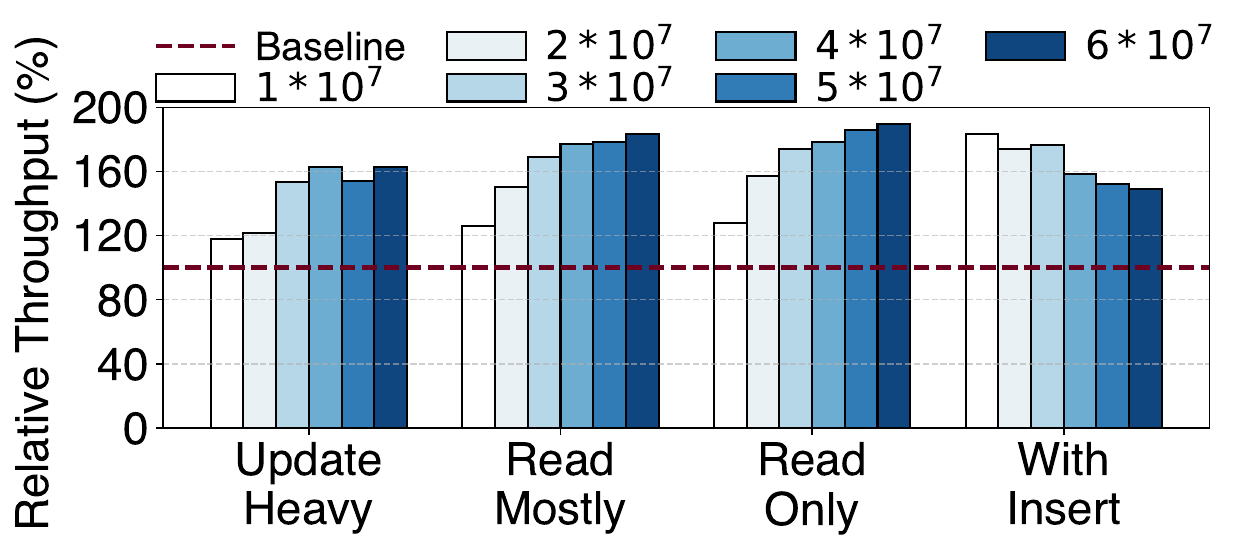}
        \vspace{-6pt}
        \label{fig:tree-size-scalability}}
     \vspace{-10pt}
    \caption{\textbf{Scalability of {\sys} ({\smass}) in microbenchmark.}}
    \label{fig:scalability}
\end{figure*}

\subsubsection{Macrobenchmark}
In the macrobenchmark, we set the baseline maximum fast memory usage at 20\% and use its corresponding memory consumption as the limit for other systems, based on Zipfian access patterns and the dataset size. 
Note that this setting with low fast memory usage is consistent with previous work~\cite{vuppalapati2024tiered,lee2023memtis,raybuck2021hemem}.
Besides this setting, we further evaluate {\sys}'s performance across various data scales and fast memory usage later to demonstrate its performance stability (\autoref{sec:diff-memory-usage}, \autoref{sec:scalability-analysis} part1).

As shown in \autoref{fig:macrobenchmark}, {\sart} outperforms the baseline by 32\%--84\%, while {\smass} outperforms the baseline by 31\%--57\% (except for \emph{Short Ranges}).
\emph{Short Ranges}'s performance depends on the number of leaves in fast memory.
Due to the limited fast memory and {\sys} prioritizing upper nodes in fast memory, the performance gain in this workload is not as significant. 
{\sys} performs better in read-dominant workloads because migration interferes less with read operations.

PAC-L performs worse than {\sart} because its static memory layout is unable to adjust to the dynamic workload.
Other data placement systems fail to perform well across all workloads. 
For instance, while TPP performs similarly to {\smass} in \emph{Update Heavy}, it fails to maintain comparable performance for \emph{Read Only} and \emph{Read Modify Write}.

\subsubsection{Latency Analysis}

We measure {\sys}'s latency under both micro and macro benchmarks. 
Due to the limited space, we only present the result of {\smass} under the microbenchmark \emph{Update Heavy} workload, though similar improvements are observed across other workloads. 
In \autoref{fig:latency}, {\sys}'s average read/write latencies are both under 1\,$\mu s$. 
Moreover, {\sys} sharply reduces read latency, with P90 latency 73\% lower than the baseline.  
{\sys} significantly reduces P90/P99 latency versus TPP and MEMTIS by continuously detecting and demoting cold nodes, along with the watermark mechanism, thereby preventing burst migrations.

\subsection{Sensitivity Analysis}
\label{sec:sensitivity-analysis}

An ideal data placement scheme should be able to adapt to changing workloads and perform well across various configurations. 
In {\sys}, most parameters are dynamically adjusted during runtime, except for workers' wake-up intervals and maximum fast memory usage. 
To evaluate {\sys}'s performance stability, we conduct a sensitivity analysis that shows
its performance under dynamic changing workloads (\autoref{sec:dynamic-workload-changes}) and different system parameter settings (\autoref{sec:diff-wakeup-interval}, \autoref{sec:diff-memory-usage}).
\subsubsection{Dynamic Workload Changes.}
\label{sec:dynamic-workload-changes}

To simulate dynamic workload changes, we shift the entire hot region in the microbenchmark every 60\,s. 
As shown in \autoref{fig:dynamic_hot_range}, although {\sys} suffers a temporary performance decline when the hot region changes---due to factors like the invalidation of hot paths and CPU cache misses---it quickly detects the new hot paths and migrates them to fast memory.
{\sys} can fully recover its performance in up to 17\,s, showing its ability to quickly adapt to changes in hot paths for stable high performance in dynamic workloads.

\subsubsection{Different Worker Wake-up Intervals.}
\label{sec:diff-wakeup-interval}
In {\sys}, the migration worker, cooler, and watermark maintainer are regularly awakened. 
The watermark maintainer's wake-up interval has minimal impact because it only adjusts some parameters during regular awakenings.
Therefore, we only evaluate the impact of wake-up intervals using five sets of different intervals for the other two workers.
\autoref{fig:diff_wakeup_interval} shows that {\sys}'s performance is relatively stable to wake-up interval choices, with less than 5\% deviations from the average.
In real-world scenarios, wake-up intervals can be adjusted based on system conditions, such as extending the intervals when CPU resources are tight to reduce overhead.

\subsubsection{Different Maximum Fast Memory Usage}
\label{sec:diff-memory-usage}

To evaluate {\sys}'s performance with different maximum fast memory usage, \autoref{fig:diff_memory_usage} compares {\sys} with the baseline in the macrobenchmark, varying maximum fast memory usage from 10\% to 80\% of the baseline's total memory.
As the fast memory usage increases, {\sys}'s performance improves steadily. 
When 80\% data is in fast memory, {\sys}'s performance is close to all data in fast memory for most workloads.
When the maximum fast memory usage exceeds 60\%, {\sys} slightly lags the baseline in \emph{Short Ranges} because the higher fast memory usage intensifies competition for CPU cache between background workers and the scan operation.
Nevertheless, when only the frontend module is enabled (red line in \autoref{fig:diff_memory_usage}), {\sys} still outperforms the baseline.

\subsection{Scalability Performance}
\label{sec:scalability-analysis}

In this section, we evaluate the scalability of {\sys} by varying the number of threads and dataset sizes. 

\noindent
\textbf{Concurrency Scalability.}
Concurrency scalability is a key metric for assessing index performance~\cite{mao2012cache,wu2019wormhole,shanny2022occualizer}.
Comparing {\sys} with the baseline (\autoref{fig:concurrency-scalability}) reveals that {\sys} not only guarantees the index's scalability but also derives amplified performance gains over the baseline with more threads.
The gains is amplified because {\sys} benefits each thread, accumulating as the thread count grows.
At 48 threads, {\sys}'s throughput is 28.6--38.1\,$\times$ that of a single thread.

\noindent
\textbf{Dataset Size Scalability.}
An important aspect in evaluating data placement schemes is their ability to maintain high performance across varying data scales~\cite{lee2023memtis,raybuck2021hemem,yan2019nimble}.
\autoref{fig:tree-size-scalability} shows the performance gains of {\sys} compared to the baseline when loading data varying from 10 to 60 million records.
In the first three workloads, {\sys}'s improvement grows with the dataset size as the longer data access paths in larger datasets emphasize the benefits of promoting hot paths.
However, for \emph{With Insert}, new nodes creates new hot paths, requiring time for {\sys} to identify and promote them, slightly reducing gains at larger scales.
Nonetheless, {\sys} outperforms the baseline by over 50\% even at larger scales.

\subsection{Performance Breakdown}
\label{sec:performance-breakdown}

This section evaluates the impact of each technique and key/value size in {\sys}. 
We first present the factor analysis (\autoref{sec:factor-analysis}), 
then evaluate the holistic management (\autoref{sec:watermark-mechanism}).

\subsubsection{Factor Analysis for {\sys}}\ 
\label{sec:factor-analysis}

\begin{figure}[t]
    \centering
    \includegraphics[width=0.9\linewidth]{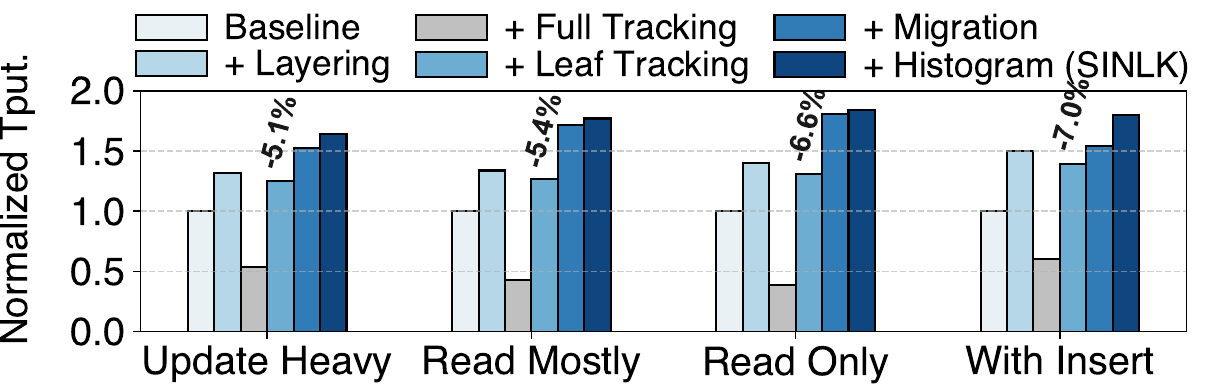}
    \vspace{-11pt}
    \caption{\textbf{The factor analysis for techniques in {\sys}.}}
    \label{fig:factors-analysis}
\end{figure}

\begin{figure}[t]
    \centering
    \vspace{-2pt}
    \begin{minipage}{0.48\linewidth}
        \centering
        \includegraphics[width=\linewidth]{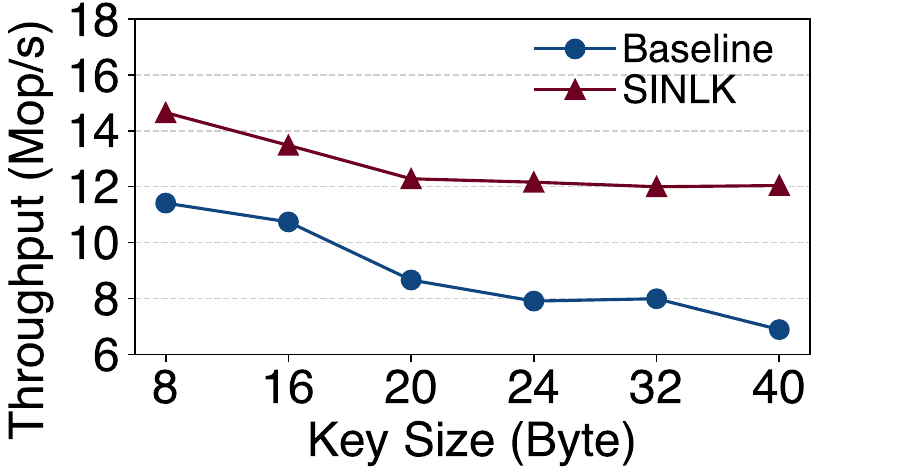}
        \vspace{-19pt}
        \caption{\emph{Impact of key sizes (Micro Update Heavy).}}
        \label{fig:vary-key-size}
        \vspace{6pt}
    \end{minipage}
    \hfill
    \begin{minipage}{0.48\linewidth}
        \centering
        \includegraphics[width=\linewidth]{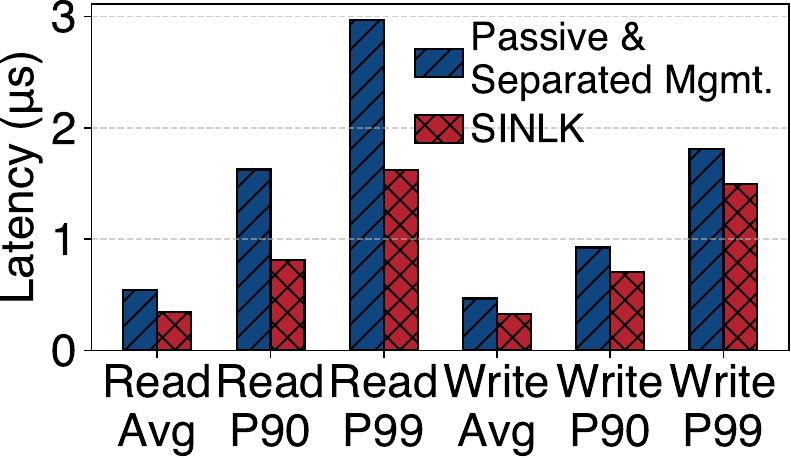}
        \vspace{-20pt}
        \caption{\emph{Impact of passive and separated management (YCSB-D).}}
        \label{fig:burst-migration}
        \vspace{-5pt}
    \end{minipage}%
    \label{fig:breakdown2}
\end{figure}

\paragraph{Contributions of techniques.}
We begin with the naive Masstree and gradually apply each proposed technique to conduct {\smass}.
\autoref{fig:factors-analysis} shows how each technique contributes to {\sys}'s performance.

\textbf{+ Layering.} 
This step implements layer-aware allocation.  
With this design, {\sys} exhibits performance gains ranging from 31.9\% to 49.8\% compared to the baseline.

\textbf{+ Access Tracking.} 
In this step, we evaluate the overhead of access tracking.
Tracking only leaf nodes incurs a minimal performance loss of 5.1\% to 7.0\% compared to the last step,
while tracking all nodes incurs about 60\% performance degradation.
This indicates that the overhead of updating only the leaf's information for access tracking is acceptable.

\textbf{+ Migration.} 
In this step, we begin to migrate hot paths and cold nodes, improving {\sys}'s performance by 2.8\% to 28.7\% over only using layer-aware allocation.
However, the fixed hot/cold thresholds are not suitable for the \emph{With Insert} workload, resulting in less noticeable performance gains.

\textbf{+ Histogram.} 
By introducing a histogram to dynamically adjust the hot/cold thresholds, {\sys} achieves performance gains ranging from 1.7\% to 16.8\% over the static thresholds.

\noindent
\paragraph{Impact of key and value size.}
We evaluate {\smass} with various key and value sizes. 
\autoref{fig:vary-key-size} shows that {\sys} achieves stable performance improvement (up to 75\%) across different key sizes.
{\sys} also stably outperforms the baseline (30\%--40\%) with value sizes from 8 to 256 bytes (we do not present the details here due to the limited space).

\subsubsection{Effectiveness of Holistic Management.}
\label{sec:watermark-mechanism}

\begin{figure}[t]
    \centering
    \includegraphics[width=0.95\linewidth]{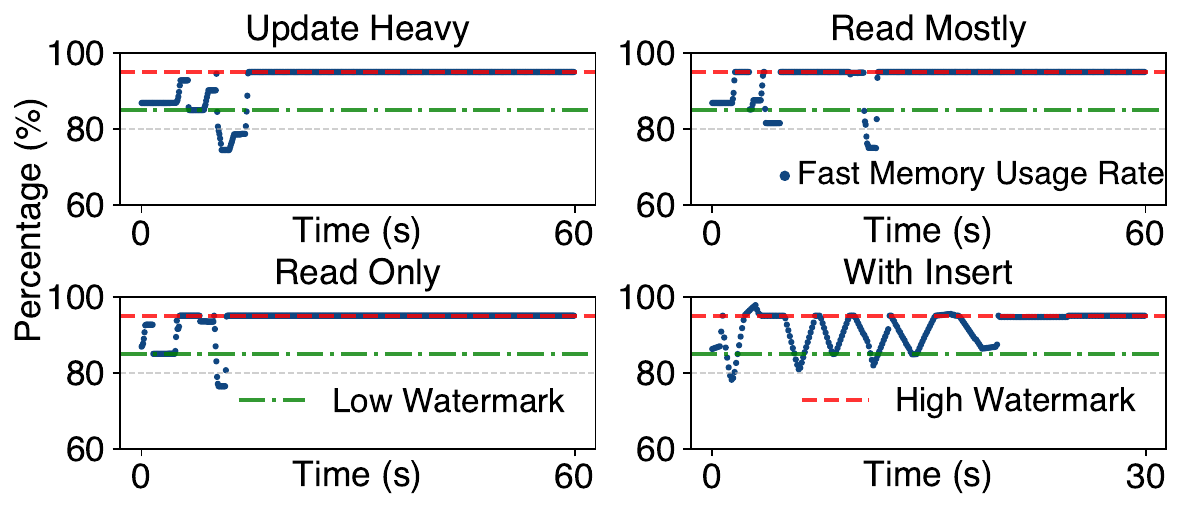}
    \vspace{-10pt}
    \caption{\textbf{Fast memory usage ratio variation over time.}}
    \label{fig:watermark-validation}
\end{figure}

\autoref{fig:burst-migration} shows that the passive (demotion only when fast memory is tight) and separated (the watermark only cares about migration) management significantly worsens the tail latency due to burst migrations when fast memory is tight.

To assess the watermark mechanism's effectiveness, we monitor the {\smass}'s fast memory usage ratio over time.
\autoref{fig:watermark-validation} shows that as hot paths are promoted, fast memory usage gradually increases. 
When exceeding the high watermark, it declines due to increased demotion.
The watermark mechanism adjusts allocation and migration to maintain fast memory usage within the watermark bounds. 
This ensures that once hot paths are promoted, fast memory usage ratio stabilizes near the high watermark across all workloads.

\subsection{Overhead Analysis}
\label{sec:overhead-analysis}

\begin{table}[t]
    \centering
    \caption{\textbf{Runtime proportion of background worker execution.} \emph{MT is migration trigger, PE/DE is promotion/demotion executor, CL is cooler, WM is watermark maintainer.}}
    \vspace{-10pt}
    \resizebox{\linewidth}{!}{
        \begin{tabular}{ccccccc}
            \toprule 
                    \textbf{Run Time Ratio (‰)} & \textbf{MT} & \textbf{PE}            & \textbf{DE}               & \textbf{CL}               & \textbf{WM} & \textbf{Total} \\ \hline
            Update Heavy                  & 1.69        & $1.37 \times 10 ^ {-2}$      & $1.12 \times 10 ^ {-2}$   & $1.04 \times 10 ^ {-2}$   & 0.799       & 2.52           \\
            Read Mostly                   & 1.52        & $2.36 \times 10 ^ {-2}$      & $8.39 \times 10 ^ {-3}$   & $3.64 \times 10 ^ {-3}$   & 0.529       & 2.09           \\
            Read Only                     & 1.13        & $6.77 \times 10 ^ {-3}$      & $9.01 \times 10 ^ {-3}$   & $4.29 \times 10 ^ {-3}$   & 0.365       & 1.51           \\
            Read Latest                   & 1.84        & $1.46 \times 10 ^ {-1}$      & $4.65 \times 10 ^ {-2}$   & $1.03 \times 10 ^ {-1}$   & 0.828       & 2.97           \\
            Short Ranges                  & 1.55        & $6.91 \times 10 ^ {-6}$      & $5.40 \times 10 ^ {-2}$   & $4.02 \times 10 ^ {-2}$   & 0.636       & 2.28           \\
            RMW                           & 1.76        & $1.01 \times 10 ^ {-1}$      & $1.44 \times 10 ^ {-2}$   & $1.06 \times 10 ^ {-2}$   & 0.839       & 2.73           \\
            \bottomrule
        \end{tabular}
    }
    \label{tab:overhead}
    \vspace{-4pt}
\end{table}

\begin{figure*}[t]
    \centering
    \subfloat[Throughput]{
        \includegraphics[width=0.14\textwidth]{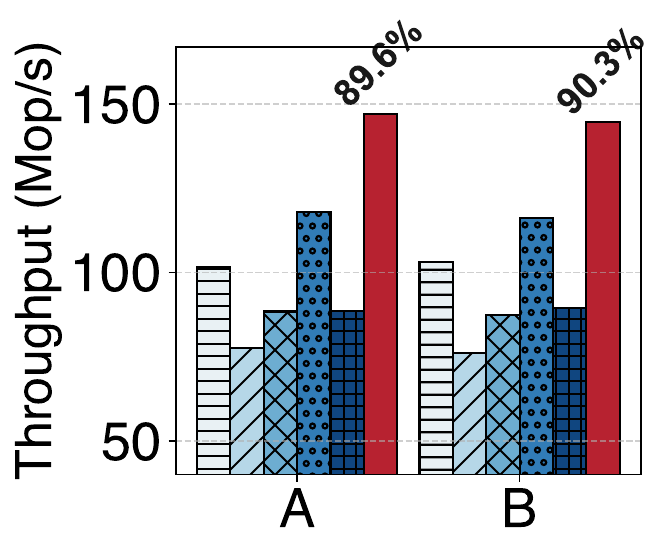}
        \vspace{-8pt}
        \label{fig:real_world_throughput_art}}
    \subfloat[Read latency]{
        \includegraphics[width=0.41\linewidth]{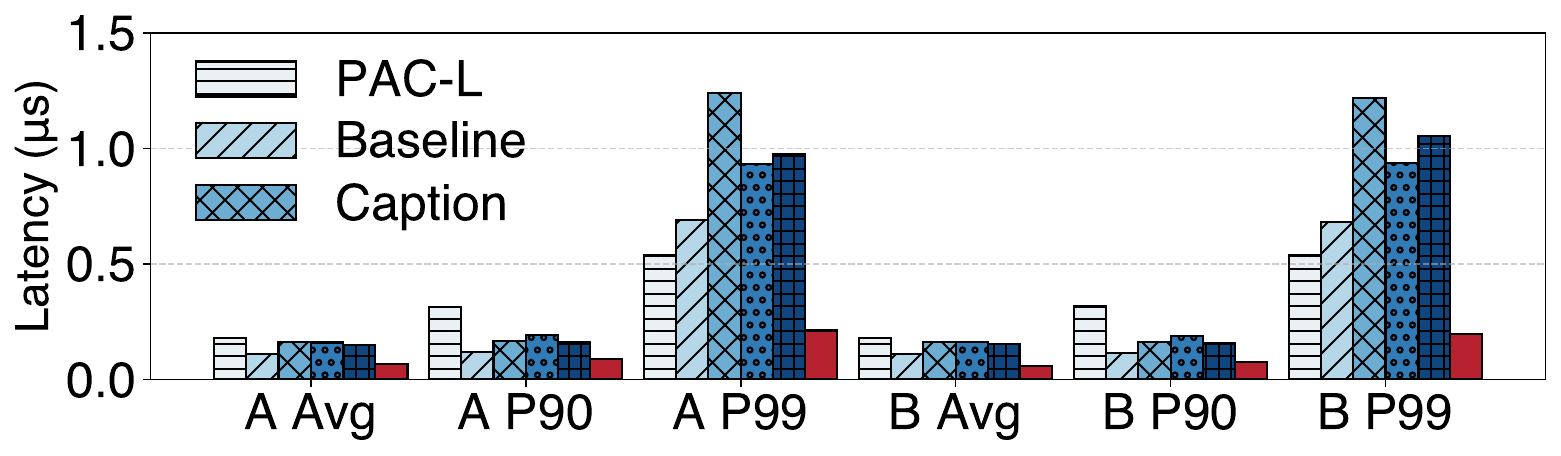}
        \vspace{-8pt}
        \label{fig:real_world_read_latency_art}}
    \subfloat[Write latency]{
        \includegraphics[width=0.41\linewidth]{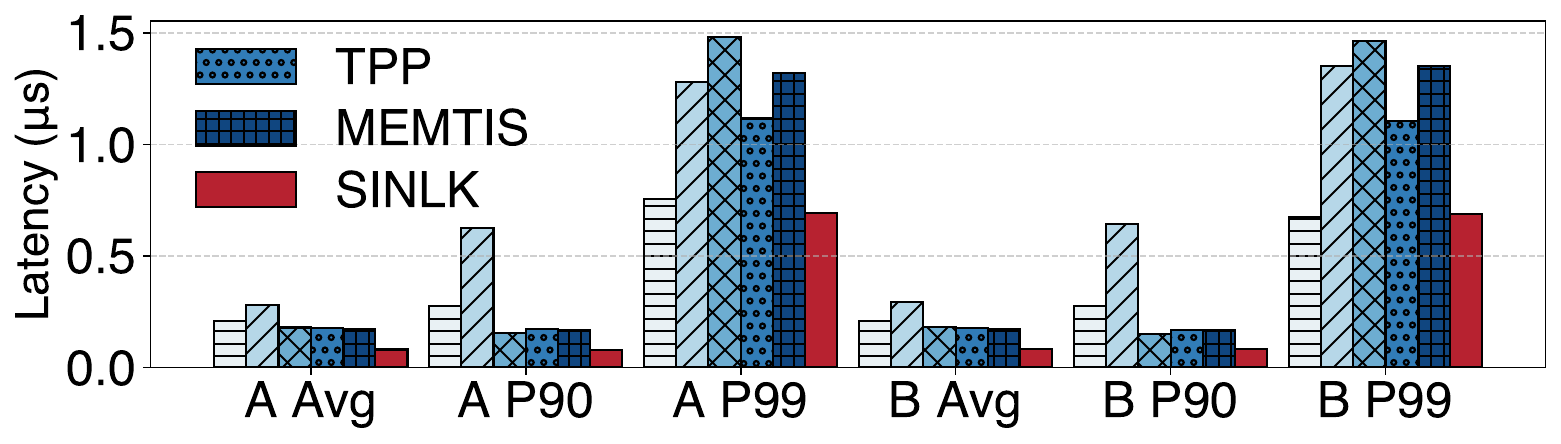}
        \vspace{-8pt}
        \label{fig:real_world_write_latency_art}}
     \vspace{-10pt}
    \caption{\textbf{Performance comparison between S-ART and compared systems with Alibaba Block Traces.}}
    \vspace{-12pt}
    \label{fig:real_world_art}
\end{figure*}

\begin{figure*}[t]
    \centering
    \subfloat[Throughput]{
        \includegraphics[width=0.14\textwidth]{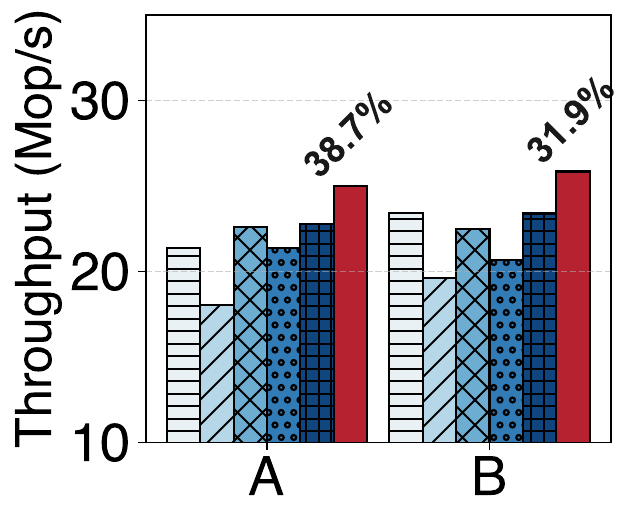}
        \vspace{-8pt}
        \label{fig:real_world_throughput}}
    \subfloat[Read latency]{
        \includegraphics[width=0.41\linewidth]{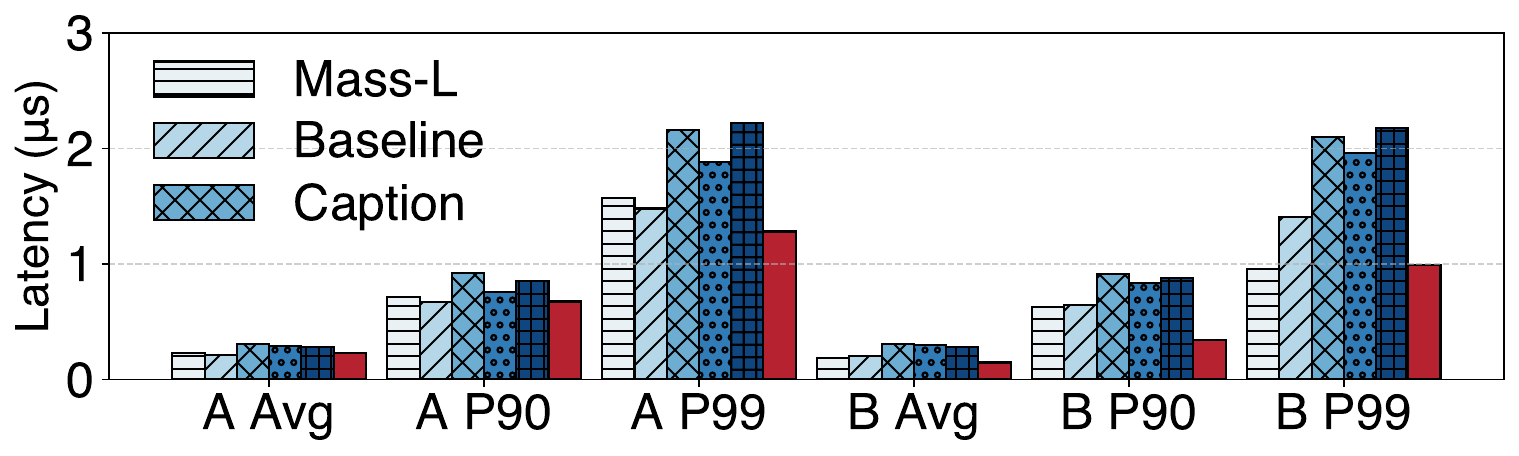}
        \vspace{-8pt}
        \label{fig:real_world_read_latency}}
    \subfloat[Write latency]{
        \includegraphics[width=0.41\linewidth]{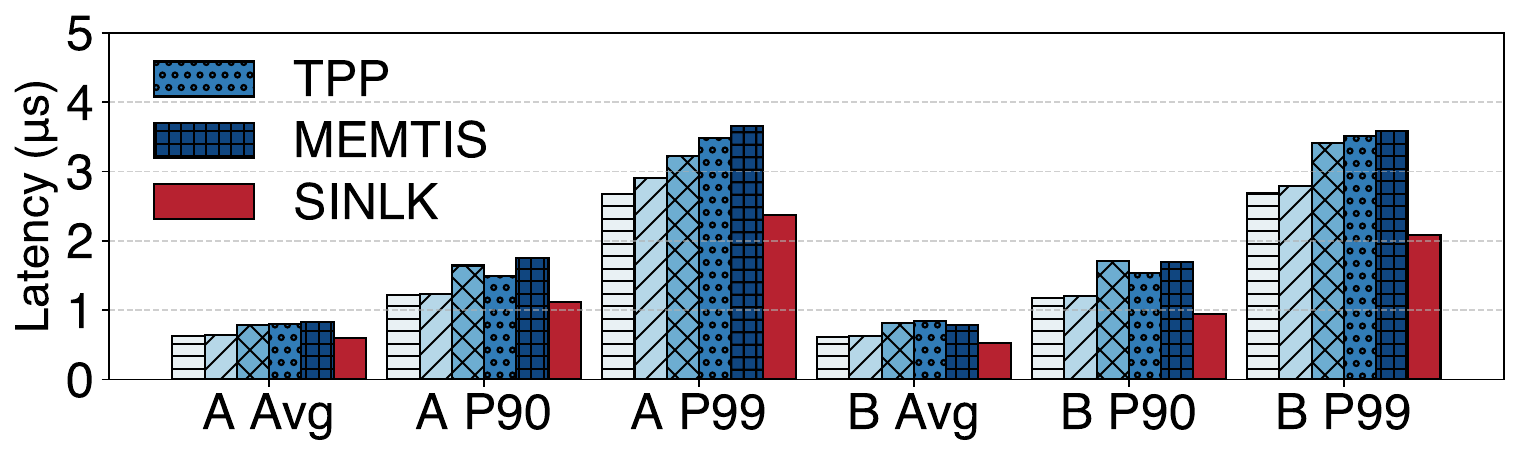}
        \vspace{-8pt}
        \label{fig:real_world_write_latency}}
     \vspace{-10pt}
    \caption{\textbf{Performance comparison between S-Masstree and compared systems with Alibaba Block Traces.}}
    \vspace{-8pt}
    \label{fig:real_world_masstree}
\end{figure*}

\noindent
\textbf{CPU Time Overhead.}
We assess the overhead of each background worker in {\sys} by measuring the proportion of its execution time to the total CPU time. 
\autoref{tab:overhead} shows that the background workers only accounts for 1.51‰--2.97‰ of the total runtime. 
The migration trigger takes the longest time among all workers, as it scans all leaf nodes to identify hot paths and cold nodes. 
Next is the watermark maintainer, which is frequently activated to monitor fast memory usage.

\noindent
\textbf{Memory Overhead.}
{\sys} only increase memory usage of Masstree's internal/leaf nodes by only 0.36\%/0.93\%.
For ART's internal/leaf nodes, the increase is 0.04\%--1.3\%/2.0\%.

\subsection{Real-world Workload Evaluation} 
\label{sec:real-world-eval}

Alibaba Block Traces, collected from Alibaba Cloud elastic block service~\cite{zhang2024whats}, reflect representative user behaviors in the cloud~\cite{wang2022separating}.
We choose two traces with the largest working set (i.e., those with the greatest need for memory expansion) for evaluation (Trace 7 as Trace A, and Trace 40 as Trace B).
\autoref{fig:real_world_art} shows that {\sart} outperforms the baseline by up to 90\% in throughput while reducing P99 latency by up to 79\% compared to MEMTIS and 63\% compared to PAC-L.
\autoref{fig:real_world_masstree} reveals that {\smass} increases throughput by up to 39\% compared to the baseline and reduces P99 latency by up to 54\% compared to MEMTIS.
This result demonstrates that {\sys} is also effective in real-world scenarios.

\section{Discussion and Future Work}%
\label{sec:discuss}

\noindent
\textbf{False Negative Internal Nodes.}
{\sys} selects promotion candidates starting from hot leaf nodes. 
In this case, a large group of infrequently accessed leaf nodes may not trigger promotion individually, but their aggregated accesses could trigger promotion of the ancestors (i.e., false negative internal nodes).
However, we find this situation is rare in practice.
We break down the accesses to a B+ tree with 20 million keys following the standard Zipfian distribution and find that false negative internal nodes only appear in the last level internal nodes with a probability of less than 0.9\%.

\noindent
\textbf{Hardware-based Access Tracking.} 
{\sys} updates the access frequency on each leaf access and traverses all leaves to select migration candidates.
The former inevitably affects tree operation performance, and the latter is the main overhead in {\sys}'s background processing (\autoref{sec:overhead-analysis}).
In the future, we will explore hardware-based sampling, like Intel Precise Event Based Sampling (PEBS)~\cite{PEBS2022Intel} and AMD Instruction-Based Sampling (IBS)~\cite{AMD2023uProfUserGuide}, to further reduce this overhead.

\noindent
\textbf{Multi-layer Memory Hierarchy.}
Currently, {\sys} targets two-layer memory hierarchies with fast and slow memory.
To extend {\sys} to multi-layer memory hierarchies, we can attempt to rank various memory types by their latency and throughput, and then apply {\sys}'s mechanisms and strategies between each pair of adjacent layers.

\section{Related Work}%
\label{sec:related}

\textbf{Page-level Data Placement.} %
Page-level data placement schemes aim to store hot pages in the fast memory and cold pages in the slow memory, while migrating pages to meet the workload demands~\cite{duraisamy2023towards,agarwal2017thermostat,lagar2019software,tiering_0.8,lee2023memtis,kim2021exploring,maruf2023tpp,xiang2024nomad,sun2023demystifying,vuppalapati2024tiered,kannan2017heteroos,ren2024mtm,xu2024flexmem}.
These schemes typically involve three major tasks: access tracking~\cite{kommareddy2019page,ryoo2018case,yan2019nimble}, hotness detection~\cite{agarwal2017thermostat,choi2021dancing,lee2023memtis}, data migration~\cite{riel2014automatic,tiering_0.8,maruf2022multi}.
{\sys} is inspired by the ideas and methods from these works, and designs the data placement scheme for tree-structure indexes on {\chm}.
Note that {\sys} does not consider memory contention, Colloid~\cite{vuppalapati2024tiered} may further enhance {\sys} when memory contention occurs. 

\noindent
\textbf{Object-granularity Data Placement. } %
The first type of user-space object-granularity data placement includes libraries~\cite{moon2023cache,cache_lib,dulloor2016data,wu2017unimem,zhang2020optimal} like CacheLib~\cite{moon2023cache,cache_lib} and X-Mem~\cite{dulloor2016data}, which require specialized APIs or offline profiling for statistics.
HeMem~\cite{raybuck2021hemem} avoids offline profiling and new APIs but does not manage the small allocation and migrates data at page granularity. 
The second type is hardware-based cacheline granularity placement~\cite{sun2025m5, zhong2024managing,lepers2023johnny}, such as Intel Flat Memory Mode and Johnny Cache.
Additionally, many databases are exploring data placement for {\chm}~\cite{lerner2024cxl,ahn2022enabling,lee2023elastic,lee2024database,huangpasha}.
Unlike these works, {\sys} leverages the tree's characteristics to achieve fine-grained management and get more semantics than hardware-based solutions without offline profiling.

\noindent
\textbf{Far Memory Management. } %
Far memory systems~\cite{yan2023patronus,marcos2018remote,ankit2020adaptive,al2020effectively,zhong2024unimem} utilize remote memory connected via network as the memory expansion.
Unlike {\chm}, far memory requires different interfaces and protocols to access remote memory, such as RDMA and RPCs.
They usually focus on reducing network overhead~\cite{chenxi2023canvas,yifan2023hermit}, read/write amplification~\cite{guo2023mira,ruan2020aifm}, and synchronization costs~\cite{shen2023ditto}.
Far memory systems and {\sys} are both related and orthogonal, as far memory indexes can benefit from {\sys}'s insights and {\sys} can leverage far memory as a new slow memory layer to extend memory.

\section{Conclusion}%
\label{sec:concl}

This paper proposes {\sys}, a fine-grained, structure-aware data placement scheme for tree-structure indexes that utilizes the tree's inherent characteristics to achieve high stable performance.

\newpage
\bibliographystyle{unsrt}
\bibliography{sinlk}

\newpage
\end{document}